\documentclass[twocolumn]{aastex631}

\usepackage{amsmath}
\usepackage{color}
\usepackage{ulem}
\usepackage{amssymb}
\usepackage{float}

\definecolor{orange}{RGB}{255,127,0}
\definecolor{grey}{RGB}{200,200,200}
\definecolor{darkgreen}{rgb}{0.1,0.5,0.1}

\newcommand{\zs}{z_{\textrm{s}}}                                                
\newcommand{\zl}{z_{\textrm{l}}}

\shorttitle{black-hole-to-halo mass relation}
\shortauthors{Li et al.}

\graphicspath{{./}{figures/}}

\begin{document}

\title{Black-Hole-to-Halo Mass Relation From UNIONS Weak Lensing}

\author[0000-0003-3616-6486]{Qinxun Li}
\affiliation{CAS Key Laboratory for Research in Galaxies and Cosmology, Department of Astronomy,\\University of Science and Technology of China, Hefei, Anhui 230026, China}
\affiliation{Department of Physics and Astronomy, University of Utah, Salt Lake City, Utah 84102, USA}

\author[0000-0001-9513-7138]{Martin Kilbinger}
\affiliation{Université Paris-Saclay, Université Paris Cité, CEA, CNRS, AIM, 91191, Gif-sur-Yvette, France}
 
\author[0000-0003-1297-6142]{Wentao Luo}
\affiliation{CAS Key Laboratory for Research in Galaxies and Cosmology, Department of Astronomy,\\University of Science and Technology of China, Hefei, Anhui 230026, China}

 \author[0000-0002-3775-0484]{Kai Wang}
\affiliation{The Kavli Institute for Astronomy and Astrophysics, Peking University, Beijing, China}
 
\author[0000-0002-4911-6990]{Huiyuan Wang}
\affiliation{CAS Key Laboratory for Research in Galaxies and Cosmology, Department of Astronomy,\\University of Science and Technology of China, Hefei, Anhui 230026, China}
\affiliation{School of Astronomy and Space Science, University of Science and Technology of China, Hefei 230026, China}

\author[0000-0002-8173-3438]{Anna Wittje}
\affiliation{Ruhr University Bochum, Faculty of Physics and Astronomy, Astronomical Institute (AIRUB), German Centre for Cosmological Lensing, 44780 Bochum, Germany}

\author[0000-0002-9814-3338]{Hendrik Hildebrandt}
\affiliation{Ruhr University Bochum, Faculty of Physics and Astronomy, Astronomical Institute (AIRUB), German Centre for Cosmological Lensing, 44780 Bochum, Germany}

\author{Ludovic van Waerbeke}
\affiliation{Department of Physics and Astronomy, University of British Columbia, Vancouver, BC V6T 1Z1, Canada}

\author[0000-0002-1437-3786]{Michael J.~Hudson}
\affiliation{Department of Physics and Astronomy, University of Waterloo, Waterloo, ON N2L 3G1, Canada}
\affiliation{Waterloo Centre for Astrophysics,
University of Waterloo, Waterloo, ON N2L 3G1, Canada}
\affiliation{Perimeter Institute for Theoretical Physics, Waterloo, ON N2L 2Y5, Canada}

\author[0000-0002-9594-9387]{Samuel Farrens}
\affiliation{Université Paris-Saclay, Université Paris Cité, CEA, CNRS, AIM, 91191, Gif-sur-Yvette, France}

\author[0000-0002-9104-314X]{Tobías I.~Liaudat}
\affiliation{IRFU, CEA, Université Paris-Saclay, F-91191, Gif-sur-Yvette, France}

\author[0009-0008-6273-8269]{Huiling Liu}
\affiliation{CAS Key Laboratory for Research in Galaxies and Cosmology, Department of Astronomy,\\University of Science and Technology of China, Hefei, Anhui 230026, China}

\author[0000-0002-9272-5978]{Ziwen Zhang}
\affiliation{CAS Key Laboratory for Research in Galaxies and Cosmology, Department of Astronomy,\\University of Science and Technology of China, Hefei, Anhui 230026, China}
\affiliation{Université Paris-Saclay, Université Paris Cité, CEA, CNRS, AIM, 91191, Gif-sur-Yvette, France}

\author{Qingqing Wang}
\affiliation{CAS Key Laboratory for Research in Galaxies and Cosmology, Department of Astronomy,\\University of Science and Technology of China, Hefei, Anhui 230026, China}

\author{Elisa Russier}
\affiliation{Université Paris-Saclay, Université Paris Cité, CEA, CNRS, AIM, 91191, Gif-sur-Yvette, France}
\affiliation{Lawrence Berkeley National Laboratory, 1 Cyclotron Road, Berkeley, CA 94720, USA}

\author{Axel Guinot}
\affiliation{Université Paris Cité, CNRS, Astroparticule et Cosmologie, F-75013 Paris, France}

\author[0000-0002-1518-0150]{Lucie Baumont}
\affiliation{Université Paris-Saclay, Université Paris Cité, CEA, CNRS, AIM, 91191, Gif-sur-Yvette, France}

\author{Fabian Hervas Peters}
\affiliation{Université Paris-Saclay, Université Paris Cité, CEA, CNRS, AIM, 91191, Gif-sur-Yvette, France}

\author{Thomas de Boer}
\affiliation{Institute for Astronomy, University of Hawaii, 2680 Woodlawn Drive, Honolulu HI 96822}

\author{Jiaqi Wang}
\affiliation{Department of Astronomy, School of Physics and Astronomy, Shanghai Jiao Tong University, Shanghai 200240, China}

\email{martin.kilbinger@cea.fr\\
wtluo@ustc.edu.cn\\
qinxunli@ustc.edu}

\begin{abstract}
This letter presents, for the first time, direct constraints on the black-hole-to-halo-mass relation using weak gravitational lensing measurements. We construct type I and type II Active Galactic Nuclei (AGNs) samples from the Sloan Digital Sky Survey (SDSS), with a mean redshift of $0.4$ $(0.1)$ for type I (type II) AGNs. This sample is cross-correlated with weak lensing shear from the Ultraviolet Near Infrared Northern Survey (UNIONS). We compute the excess surface mass density of the halos associated with $36,181$ AGNs from $94,308,561$ lensed galaxies and fit the halo mass in bins of black-hole mass. We find that more massive AGNs reside in more massive halos. We see no evidence of dependence on AGN type or redshift in the black-hole-to-halo-mass relationship when systematic errors in the measured black-hole masses are included.
Our results are consistent with previous measurements for non-AGN galaxies.
At a fixed black-hole mass, our weak-lensing halo masses are consistent with galaxy rotation curves, but significantly lower than galaxy clustering measurements. 
Finally, our results are broadly consistent with state-of-the-art hydro-dynamical cosmological simulations, providing a new constraint for black-hole masses in simulations.

\end{abstract}


\section{Introduction} \label{sec:intro}

Supermassive black holes (SMBHs), with typical masses of $10^6-10^{10} M_{\odot}$, are among the most mysterious objects in the Universe. 
It is widely accepted that most galaxies have an SMBH in their center \citep{Kormendy1995}. 
Though the formation and evolution of SMBHs remain unclear, there is already a large amount of evidence indicating a coevolution between SMBHs and their host galaxies, \cite[see for a review][]{kormendyCoevolutionNotSupermassive2013}. 
In addition, galaxy properties are expected and have been shown to be closely related to their host dark-matter halos, as this is where they form and evolve \citep[e.g.]{doi:10.1146/annurev-astro-081817-051756}. These observational results suggest that a close connection between halos, galaxies, and SMBHs needs to be established to understand the coevolution of these different classes of objects \citep{TRINITYI, TRINITYIII}. 
The gravitational potential of a halo determines the accretion of baryons and star formation of galaxies into the halo. 
Several mechanisms in galaxies, such as bar instabilities, conduct cold gas into galaxy centers, feeding the accretion of supermassive black holes.
The energetic feedback of the accretion can push
baryons outside the galaxy or even the halo, which will suppress the
SMBH growth and star formation. Such complex interplay among halos, galaxies, and SMBHs plays a crucial role in galaxy formation and evolution and is still under exploration.

The first step towards understanding the connection between halos, galaxies, and SMBHs is to build statistical relationships between these three types of objects based on observational data. Much effort has been
devoted to this aspect in previous decades. The pioneering work was initiated by
\citet{dresslerStellarDynamicsNuclei1988}, who noted a positive correlation
between the black-hole mass and the spheroid luminosity. Subsequent studies with more extensive data sets found a tight
correlation between black-hole mass $M_{\bullet}$ and various galaxy properties, such as bulge mass and stellar velocity dispersion, across several orders of
magnitude \citep{1998AJ....115.2285M,Ferrarese2000,Gebhardt2000,Kormendy&Ho, Saglia2016}. 
There are also many studies on the galaxy-halo scaling relations, with the
stellar mass-halo mass relation, see \cite{2008ApJ...676..248Y} as a representative example.

The relation between SMBHs and their host halos has yet to be extensively studied. 
\citet{Ferrarese2002} used the maximum rotational velocity of late-type galaxies, $v_{\rm c}$, as a tracer of the halo mass, and the central velocity dispersion, $\sigma_*$, of their bulges as a tracer of the black-hole mass. This led to the first measurement of the $M_{\bullet}$-$M_{\rm h}$ relation.
This relation for galaxies was further confirmed in larger samples 
\citep{Baes2003,Pizzella2005,VNG2011} with a similar method, which used $\sigma_*$ and $v_{\rm c}$ as tracers of black-hole and halo mass.
\cite{Sabra2015}, \cite{DGC2019}, and \cite{Marasco2021} used direct dynamical black-hole mass instead of $\sigma_*$ and found a correlation between SMBH mass and dynamical halo mass. These works based on dynamics are limited to small galaxy samples and rely on strong assumptions about the kinematic state of the gas, and the density profile of the dark matter halo. 
There is also evidence for the opposite idea that the SMBH mass does not correlate with halo mass, as seen in bulgeless galaxies \citep{K&B2011}. 

Unlike quiescent SMBHs in normal galaxies, Active Galatic Nuclei (AGN) are SMBHs that are actively accreting matter. The trigger, growth, and feedback of AGNs are critical issues in the halo-galaxy-SMBH connection. For the $M_{\bullet}$-$M_{\rm h}$ relation in AGN samples, the halo mass in different bins of $M_{\bullet}$ is typically inferred from the spatial two-point correlation function of AGNs together with empirical models such as Halo Occupation Distribution (HOD) and abundance matching \cite[e.g.][]{Krumpe2015, Powell2018, Shankar2020, Powell2022,krumpe2023spatial}. Using gas dynamics, the $M_{\bullet}$-$M_{\rm h}$ relation has been measured from $z=0$ to $z=6$ using reverberation-mapping and virial black-hole masses \citep{Robinson2021, Shimasaku2019}. More massive AGNs were more likely found in more massive halos. However, the methods used to estimate the halo mass in the works listed above are indirect and strongly model-dependent.

Gravitational lensing, an effect directly related to the density field, has been emerging as the most direct and clean method to measure halo mass \citep{2006MNRAS.368..715M,2018ApJ...862....4L}. For galaxies, \cite{Bandara2009} and \cite{Zhang2023} inferred black-hole masses from the $M_{\bullet}$-$\sigma_*$ relation, and measured halo masses with strong lensing and weak lensing, respectively. Their results significantly differ from the $M_{\bullet}$-$M_{\rm h}$ relation from AGN clustering. Previous weak-lensing studies on AGNs focused on the $M_*$-$M_{\rm h}$ relation, using samples with limited size \citep{2009MNRAS.393..377M,2015MNRAS.446.1874L,2022arXiv220403817L}. In this work, we use, for the first time, weak lensing to constrain the AGN $M_{\bullet}$-$M_{\rm h}$ relation. We utilize the Sloan Digital Sky Survey (SDSS) AGN sample together with the galaxy shape catalog derived from the Ultraviolet Near Infrared Optical Northern Survey (UNIONS) imaging data, achieving a high signal-to-noise ratio measurement of the $M_{\bullet}$-$M_{\rm h}$ relation. 

This paper is organized as follows: Sect.~\ref{sec:data} introduces the AGN lens samples and weak-lensing galaxy shape catalogs, Sect.~\ref{sec:methods} presents our methodology, before we show and discuss our results in Sect.~\ref{sec:result}. 

Throughout this work, we assume a Planck18 cosmology \citep{Planck18}.

\section{Data} \label{sec:data}

\subsection{Lens sample}

In this work, we construct type I and type II AGN/quasar samples as our lens samples based on three SDSS spectroscopic cataloguess, described in the following sections.

\begin{figure}
    \centering
    \includegraphics[width=1\columnwidth]{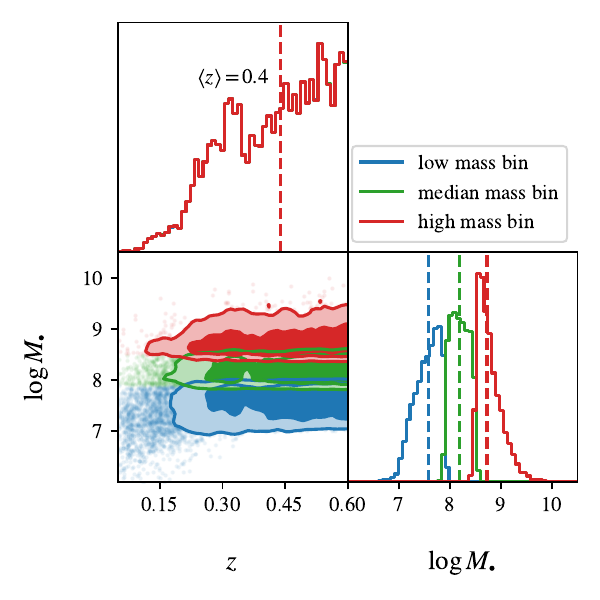}
    
    \caption{Joint and marginalised redshift and black-hole mass distribution of the type I samples. Low, median, and high black-hole mass bins are shown in blue, green, and red, respectively. We use the redshift-distribution weights $w_{\rm l, nofz}$ and the lensing efficiency weights $(\overline{\Sigma^{-1}_{\textrm{cr}}})^2$ for the 1D and 2D distributions. }
    \label{fig:dis}
\end{figure}

\subsubsection{SDSS type I AGNs}

Based on the SDSS DR16 quasar catalog \citep{Lyke2020}, Wu \& Shen fitted the spectrum of $750,414$ quasars in the redshift range $0.1<z<6$ and measured virial black-hole masses. As an update to \cite{Shen11}, they used the FWHM of H$\beta$, Mg II, and C IV broad emission lines (combined with the broad-line-region radius inferred from continuum luminosity) for their estimates. Here, we adopt their black-hole masses based on $H\beta$. The mean statistical error in $\log M_{\bullet}$ is much smaller than the systematic error of the virial black-hole mass \cite[$\sim0.4$ dex, see][]{Shen2013}.

As a complement to \cite{Wu&Shen2022} at low black-hole masses, we use the AGN catalog from \cite{Liu2019}, a complete AGN sample including both quasars and Seyfert galaxies from SDSS DR7. Black-hole masses are measured with H$\alpha$ and H$\beta$, and we adopt the $H\beta$ mass. This catalog contains $14,584$ AGNs at $z<0.35$.

We merge the two catalogs and remove duplicate objects. As a consistency check, we compare the fiducial black-hole mass of duplicate objects from both catalogs and find no significant systematic bias.
In this work, we use AGNs with redshifts $0.05<z<0.6$ in the overlapping sky region between SDSS and UNIONS, resulting in $14,649$ lenses, three times larger than the sample size of previous type I AGN weak lensing studies \citep{2022arXiv220403817L}. We divide the sample into low ($\log M_{\bullet}/M_{\odot}<7.9$), median ($7.9<\log M_{\bullet}/M_{\odot}<8.5$) and high ($\log M_{\bullet}/M_{\odot}>8.5$) black-hole mass bins. We introduce a weight $w_{\rm l,nofz}$ such that the weighted redshift distributions of the low and median mass bins equal the high-mass bin. This allows for a fair comparison between the mass bins free of redshift evolution or selection biases. The weighted distributions of the three bins are shown in Fig.~\ref{fig:dis}. 

\subsubsection{SDSS type II AGNs}

In addition to the type I catalog, we construct a type II AGN sample from the SDSS DR7 MPA-JHU catalog \citep{2003MNRAS.341...33K,2004MNRAS.351.1151B}. We identify galaxies classified as AGNs using the BPT diagram \citep{1981PASP...93....5B} within the catalog. We estimate black-hole masses from the velocity dispersion using the $M_{\bullet}$-$\sigma_*$ relation proposed by \cite{Saglia2016}. To correct for the aperture effect of velocity dispersion, we adopt the method outlined by \cite{2006MNRAS.366.1126C}: First, we cross-match the sample with the NYU-VAGC catalog \citep{2005AJ....129.2562B} to obtain the $r$-band effective radius $R_{\rm e}$. Subsequently, we compute the aperture-corrected velocity dispersion, $\sigma_*$, using the formula $\sigma_*=\sigma_{\rm ap}(R_{\rm e}/R_{\rm ap}/8)^{-0.066}$, where $\sigma_{\rm ap}$ is the fiber velocity dispersion, and $R_{\rm ap}=3\arcsec$ is the fiber aperture for SDSS spectra. Finally, we restrict the sample to $21,532$ AGNs within the redshift range $z\in[0,0.2]$, black-hole mass range $\log M_{\bullet}/M_\odot \in [6.67,9.33]$, and falling in the UNIONS footprint. The type II sample exhibits lower redshifts ($\langle z\rangle\simeq 0.1$) than the type I sample.

\subsection{Source sample}
\label{sec:source_sample}

The shape catalogs serving as the background source sample in this work are the v1.3 ShapePipe and v1.0 \textit{lens}fit catalogs of UNIONS\footnote{https://www.skysurvey.cc/}. UNIONS is an ongoing multi-band wide-field imaging survey conducted with three telescopes (Canada-France-Hawai'i Telescope for $u$ and $r$ bands, Subaru telescope for $g$ and $z$ bands, and Pan-STARRS for the $i$ band) in Hawai'i. UNIONS will cover $4,800$ square degrees of the Northern sky with deep exposures and high-quality images. The depth (limiting magnitude with point source 5-sigma in a $2\arcsec$ diameter aperture) reaches $24.3$, $25.2$, $24.9$, $24.3$, and $24.1$ in $u, r, g, i$ and $z$, respectively.

At the time when the shear catalogs were produced (beginning of 2022), the survey covered an area of around $3,500$ square degrees in the $r$-band (the galaxy shapes were measured in this band). 
We did not have photometric redshifts for each source galaxy in the catalog at this stage of the UNIONS processing since the observations and calibration of the multi-band photometry are still ongoing. Instead, we estimated the overall redshift distribution by a method based on self-organizing maps. See Appendix~\ref{sec:photozdetail} for more details about photometric redshifts.

The ShapePipe catalog was processed with the \textsc{ShapePipe} software package \citep{2022A&A...664A.141F}. It contains $98$ million galaxies over an area of $3,200$ square degrees effective area. An earlier version of the ShapePipe catalog was published in \cite{2022A&A...666A.162G}. Some updates in processing were implemented for the v1.3 shear catalog used here, as follows. First, to model the PSF, instead of PSFEx \citep{2011ASPC..442..435B} we used MCCD \citep{MCCD21} that performs a non-parametric Multi-CCD fit of the PSF over the entire focal plane. Second, we reduced the minimum area to detect an object from $10$ to $3$ pixels via the \textsc{SExtractor} configuration keyword \texttt{DETECT\_MINAREA = 3}. This leads to a smaller galaxy selection bias on the ensemble shear estimates. Third, we added the section on the relative size between galaxies, $T_\textrm{gal}$, and the PSF, $T_\textrm{PSF}$, as $T_\textrm{gal} / T_\textrm{PSF} < 3$ to avoid contamination by very diffuse, mostly low-signal-to-noise objects which tend to be artefacts.

The \textit{lens}fit shape catalog was created with the THELI processing and \textit{lens}fit software \citep{2007MNRAS.382..315M}. It contains $109$ million galaxies in a $2,100$ square degree sky area.
The effective area and number density of the ShapePipe and \textit{lens}fit catalogs are different due to masking and processing choices. In the following, we use the more conservative \textit{lens}fit mask for both shape catalogs, defining the common UNIONS footprint in which SDSS AGNs are selected. Both \textit{lens}fit and ShapePipe catalogs are based on the same image data.

\section{Methods}
\label{sec:methods}

\begin{figure*}[bht]
    \centering
\includegraphics[width=2\columnwidth]{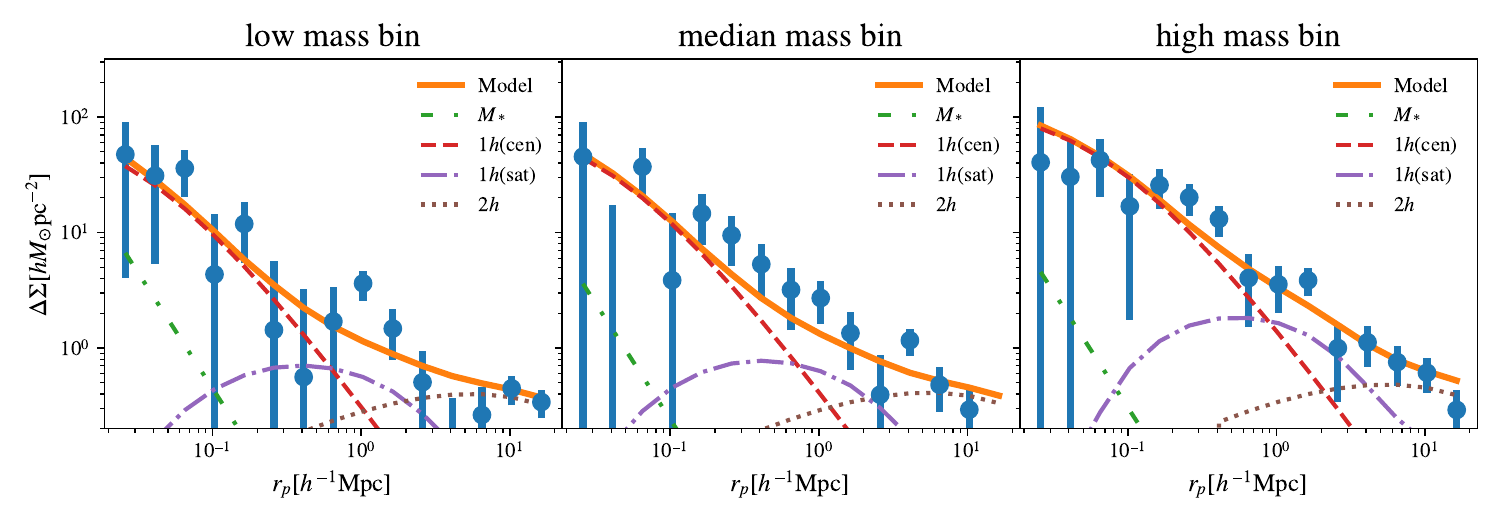}
    
    \caption{Galaxy-galaxy lensing excess surface mass density of three black-hole mass bins from ShapePipe. The best-fit HOD models are presented in orange lines. The baryon contribution, one-halo term of centrals, one-halo term of satellites, and two-halo term are plotted in green, red dashed, purple dash-dotted, and brown dotted lines, respectively. The measurements from the \textit{lens}fit catalog are similar. }
    \label{fig:signal}
\end{figure*}

\subsection{Galaxy-galaxy lensing technique}
\label{sec:ggl_basics}

Galaxy-galaxy lensing denotes the shape distortions of background source galaxies due to the gravitational field of matter associated with foreground lens galaxies \cite[see for a review][]{K15}. The main \emph{physical} quantity related to galaxy-galaxy lensing is the excess surface density (ESD), $\Delta \Sigma$, at a projected distance $R$, defined as the mean surface density within a disk of radius $R$ minus a boundary term, which is the mean surface mass at radius $R$,
\begin{equation}         
    \Delta \Sigma(R) = \bar \Sigma(< R) - \Sigma(R).
\end{equation}
The main \emph{observable} for galaxy-galaxy lensing is the tangential shear, $\gamma_\textrm{t}$, of a source sample induced by a lens at projected distance $R$. This observable is related to the ESD via
\begin{equation} 
    \Delta \Sigma(R) = \Sigma_{\textrm{cr}} \gamma_\textrm{t}(R),
    \label{eq:Delta_Sigma}
\end{equation} 
where the critical surface mass density $\Sigma_\textrm{cr}$ is defined as            
\begin{equation}
    \Sigma_{\textrm{cr}}(\zl, \zs) = \frac{c^2}{4 \pi G}     
        \frac{ d(\zs)  }{ d(\zl) \, d(\zl, \zs) }\frac{1}{(1+\zl)^2} .    
    \label{eq:Sigma_cr}  
\end{equation}     
Here, $\zs$ ($\zl$) is the source (lens) redshift, and
$d(\zs)$,  $d(\zl)$, and  $d(\zl, \zs)$ are the angular diameter distance from the observer to the source, to the lens, and the lens-source distance, respectively.
The constants are the speed of light $c$ and the Newtonian gravitational constant $G$.

\subsection{Estimators}
\label{sec:estimators }

An estimator for the tangential shear of a background source sample around a lens galaxy population is
\begin{equation}
    \left\langle \gamma_\textrm{t}(R) \right\rangle
        = \frac{\sum_{\rm l s} w_{\rm l} w_{\rm s} \epsilon_{\rm t,s}
                1_{b(R)}(| \vec r_{\rm l} - \vec r_{\rm s}|)
                }{ \sum_{\rm l s} w_{\rm l} w_{\rm s} } .   
    \label{eq:gamma_t_est}   
\end{equation}
This estimator is a weighted sum over the observed tangential ellipticities, $\epsilon_{\rm t,s}$, of source galaxies around lens galaxies. Source galaxies have a weight, $w_{\rm s}$, stemming from the galaxy shape estimation that indicates measurement uncertainties. Lens weights, $w_{\rm l} \equiv w_{\rm l, nofz}$, are introduced to homogenize the redshift distribution across lens samples as discussed in Sect.~\ref{sec:source_sample}. The indicator function $1_S(x)$ of the set $S$ is unity if $x \in S$ and zero otherwise. In the above equation, this function selects galaxy pairs in a bin $b(R)$ around the projected separation $R$ of the pair; the shape of the function $b$ is chosen to be logarithmic.

Since we do not have photometric redshifts of individual background galaxies in the shape catalog,
we compute an effective surface mass density by averaging Eq. \eqref{eq:Sigma_cr} over the source redshift distribution.
Inserting this effective value into
Eq.~\eqref{eq:Delta_Sigma} results in an average excess surface mass density. Since we cannot select sources to be strictly behind the lens sample, this leads to a divergence of $\Sigma_{\rm cr}$ when $\zs\to\zl$. A practical solution is to compute the inverse effective critical surface mass density,
\begin{equation}
    \overline{\Sigma^{-1}_{\textrm{cr}}}(\zl) = \frac{4 \pi G}{c^2} \, d(\zl)(1+\zl^2) \,
        \int_0^{z_\textrm{lim}} \textrm{d} \zs \, n(\zs) \, \frac{ \, d(\zl, \zs) }{ d(\zs) } .
    \label{eq:Sigma_cr_eff_inv}
\end{equation}
This quantity is the inverse of the critical surface mass density $\Sigma_\textrm{cr}$, Eq.~\eqref{eq:Sigma_cr}, weighted by the source redshift distribution (see Sect.~\ref{sec:data}).
The effective excess surface mass density is then
\begin{equation}            
    \overline{\Delta \Sigma}(R) = \gamma_\textrm{t}  
    \left[ \overline{\Sigma^{-1}_{\textrm{cr}}}(\zl) \right]^{-1} .
\end{equation}
Using Eq.~\eqref{eq:Sigma_cr_eff_inv}, a first estimator for the excess surface mass density is readily derived as
\begin{equation} 
    \left\langle \Delta \Sigma(R) \right\rangle^\prime    
        = \frac{\sum_{\rm l s} w_{\rm l} w_{\rm s} \epsilon_{\rm t,s}   
        \left[ \overline{\Sigma^{-1}_{\textrm{cr}}}(\zl) \right]^{-1}           
                1_{b(R)}(| \vec r_{\rm l} - \vec r_{\rm s}|)             
                }{ \sum_{\rm l s} w_{\rm l} w_{\rm s} } .        
    \label{eq:Delta_Sigma_est}
\end{equation}
When using the effective surface mass density, the weights for a given lens can be updated by multiplication with the square of the inverse effective critical surface mass density Eq.~\eqref{eq:Sigma_cr}, to down-weigh lenses with a low lensing efficiency, 
$w_{l} \to w_{\rm l} \left( \overline{\Sigma_\textrm{cr}^{-1}}(z_{\rm l}) \right)^2$. With this, we write our final estimator of the excess surface mass density as
\begin{align}
\left\langle \Delta \Sigma(R) \right\rangle
      = \frac{\sum_{\rm l s} w_{\rm l} \,
        \overline{\Sigma_\textrm{cr}^{-1}}(z_{\rm l})
        \, w_{\rm s} \, \epsilon_{\rm t,s}
                1_{b(R)}(| \vec r_{\rm l} - \vec r_{\rm s}|)
                }{ \sum_{\rm l s} w_{\rm l} w_{\rm s} \left( \overline{\Sigma_\textrm{cr}^{-1}}(z_{\rm l}) \right)^2 } .
    \label{eq:Delta_Sigma_est_2}
\end{align}
We also conduct a series of systematic tests and apply the boost factor correction $\left\langle \Delta \Sigma(R)^\textrm{c} \right\rangle=B(R)\left\langle \Delta \Sigma(R) \right\rangle$. We refer to Appendix~\ref{sec:systematic} for details.

\subsection{AGN lens model} \label{sec:model}

Our lens sample contains both central and satellite AGN host galaxies, and we need to consider contributions to the excess surface density from both. We adopt a HOD model from \cite{GS02} to describe the average ESD around the AGN sample,
\begin{equation}
    \Delta\Sigma = \Delta\Sigma_{\rm b} + (1-f_{\rm sat})\Delta\Sigma_{\rm h, cen}+f_{\rm sat}\Delta\Sigma_{\rm h, sat}+ \Delta\Sigma_{2\mathrm{h}},
    \label{eq:model}
\end{equation}
where $f_{\rm sat}$ is the satellite galaxy fraction of the sample, left as a free parameter. $\Delta\Sigma_{\rm b}$ is the contribution from baryons in the host galaxy, containing the stellar mass $M_*$. $\Delta\Sigma_{\rm h, cen}$ and $\Delta\Sigma_{\rm h, sat}$ are the one-halo terms of the central and satellite galaxy, respectively. $\Delta\Sigma_{2\mathrm{h}}$ is the two-halo term. The terms are described in Appendix~\ref{sec:modeldetail} in detail. 

\section{Results and Discussion}
\label{sec:result}

\begin{figure*}
    \centering
    \includegraphics[width=2\columnwidth]{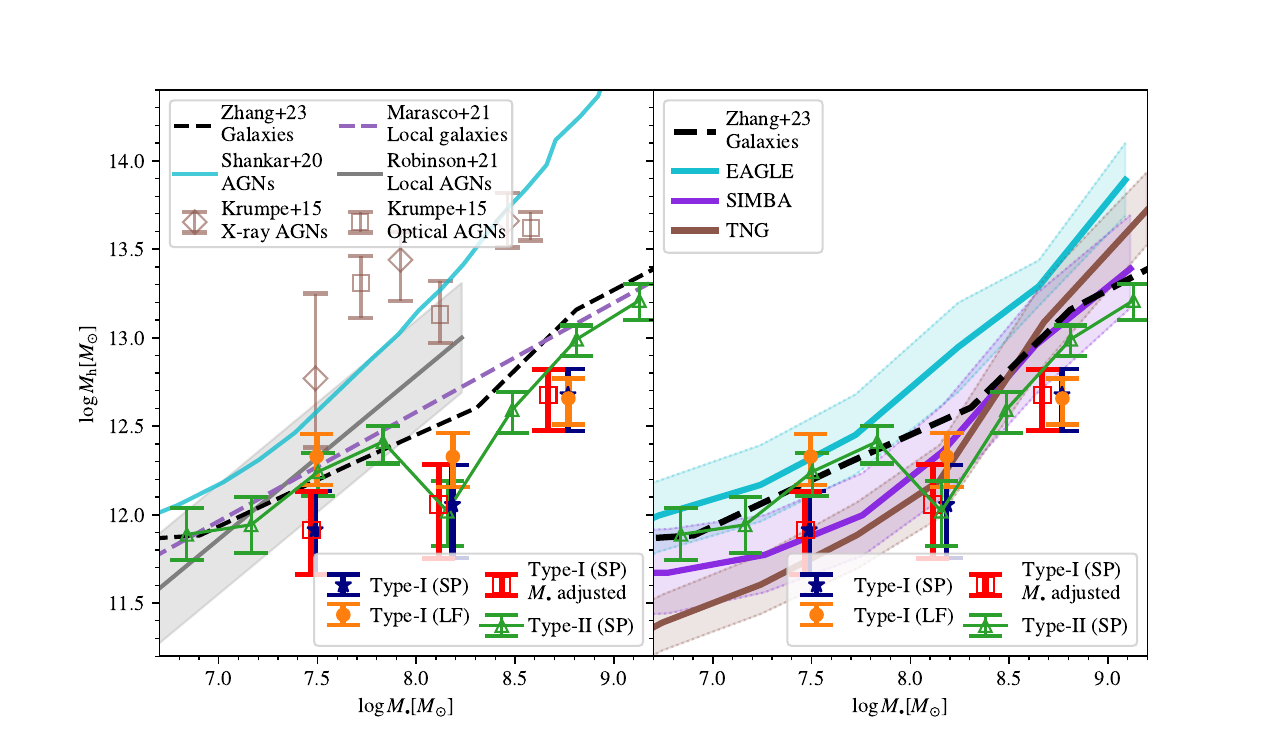}
    
    \caption{Black-Hole-to-Halo Mass Relation. In both panels, the dark blue stars and orange-filled circles are our type I results from the ShapePipe and \textit{lens}fit catalogs, respectively. The blank red squares are type I ShapePipe results with an adjustment on BH mass. Blank green triangles correspond to the type II results with ShapePipe. \textit{Left Panel}: Comparison with observational results from the literature. Brown squares and rhombuses are AGN clustering results for optically selected and X-ray selected samples from \cite{Krumpe2015}, respectively. Cyan and grey lines show AGN clustering and dynamics results from \cite{Shankar2020} and \cite{Robinson2021}, respectively. Results for normal galaxies are shown as dashed lines (purple line from \cite{Marasco2021} and black line from \cite{Zhang2023}). \textit{Right Panel}: Results from EAGLE, TNG, and SIMBA hydro-simulations are shown in cyan, brown, and purple lines with the $16-84$ percentiles displayed as shaded areas. }
    \label{fig:main_res}
\end{figure*}

\subsection{AGN black-hole-to-halo mass relation}
\label{sec:res_obs}

The measured galaxy-galaxy lensing ESD profiles for the type I sample are shown in Fig.~\ref{fig:signal}. We measure the ESD with a high signal-to-noise ratio (SNR) from ShapePipe (\textit{lens}fit), with values of $24 $ ($34$), $26$ ($36$), and $31$ ($41$) in the three bins, respectively. Our measurements are well reproduced by the HOD models with three free parameters. The amplitude of the ESD increases with $M_{\bullet}$, indicating that more massive SMBHs are situated in more massive halos. A similar trend is observed in the type II sample.

Fig.~\ref{fig:main_res} shows the $M_\bullet$ - $M_\textrm{h}$ relation. Systematic errors in the shape measurements contribute to the total error budget at a comparable level to statistical errors, as indicated by the disparity between the type I ShapePipe and \textit{lens}fit results. This underscores the robustness of our analysis across different shape catalogs.

We observe that more massive AGNs inhabit larger dark-matter halos, consistent with previous findings from dynamical \citep{Robinson2021} and clustering analyses \citep{Krumpe2015, Shankar2020}. \cite{Robinson2021} employed reverberation mapping to determine black-hole masses and utilized HI FWHM to estimate halo masses for $24$ local AGNs. 
Our results are consistent with \cite{Robinson2021} at low masses but at $\log M_{\bullet}/M_{\odot} \simeq 8$, we find a lower halo mass. 
Our measured halo masses in the high-mass regime for both type I and type II samples are also systematically lower than the clustering results reported in \cite{Krumpe2015} and \cite{Shankar2020}. The discrepancy between clustering and lensing halo masses was noticed in previous works \citep{2009MNRAS.393..377M}. The clustering method leverages the monotonic relation between halo mass and halo bias, while weak lensing directly probes the matter overdensity around the tracers. It has long been known that the clustering strength of halos also depends on their secondary properties, such as the halo structure and the halo assembly history, which is called the halo assembly bias or the secondary halo bias \citep{2007MNRAS.377L...5G, 2024MNRAS.528.2046W}. Therefore, if the host galaxies of AGNs prefer to live in dark-matter halos with biased secondary properties, it will alter the clustering strength without changing the host halo mass, while lensing is free of this effect. 

In our results,  both type I and type II samples exhibit similar $M_{\bullet}$-$M_{\rm h}$ relations despite their distinct classifications and redshift ranges. At higher black-hole masses, type I AGNs have a lower halo mass compared to type II. This difference may be interpreted as a systematic error in virial black-hole mass estimation. Recent spectroscopic interferometer measurements \citep{2024arXiv240107676G} of the AGN size-luminosity relation, upon which the virial mass measurement of type I AGNs relies, exhibit a significantly lower slope than the one proposed by \cite{2006ApJ...644..133B} from reverberation mapping. Our type I black-hole masses are calculated by \cite{Wu&Shen2022} using the H$\beta$ prescription from \cite{2006ApJ...641..689V}, which is based on \cite{2006ApJ...644..133B}. To account for the new measurements of \citep{2024arXiv240107676G}, we use their size-luminosity slope to derive a black-hole mass prescription with the same sample and the same method as \cite{2006ApJ...641..689V} and get the updated relation
\begin{equation}
\begin{split}
        \log \left(\frac{M_{\bullet}}{M_{\odot}}\right)=6.88&+2\log \left(\frac{\text{FWHM}_{\text{H}\beta}}{1000 \, \text{km s}^{-1}}\right)\\
        &+0.30\log\left(\frac{\lambda L_{\text{5100}}}{10^{44} \, \text{erg s}^{-1}}\right), 
\end{split}
\end{equation}
where $\log{\lambda L_{\text{5100}}}$ is the luminosity of the continuum at $5100\text{\AA}$.
We then adjust the average black-hole mass of the type I sample, as shown in Fig.~\ref{fig:main_res}. The black-hole mass of the high-mass bin changes the most, and moves the black-hole-halo-mass relation closer to the type II line. In conclusion, we find no evidence of type- or redshift-dependence in the $M_{\bullet}$-$M_{\rm h}$ relationship.

Furthermore, we compare our results to those of normal (non-AGN) galaxies.
\cite{Marasco2021} measured halo masses through
globular cluster dynamics and galaxy rotation curves in 55 nearby galaxies with directly measured black-hole masses. \cite{Zhang2023} used the Dark Energy Camera Legacy Survey \cite[DECaLS]{2019AJ....157..168D} shape catalog  \citep{2022AJ....164..128Z} to measure galaxy-galaxy lensing of quiescent galaxies for $z<0.2$ for different $\sigma_*$ bins. We plot their result with black-hole masses inferred from the $M_{\bullet}$-$\sigma_*$ relation of \cite{Saglia2016}. Compared to these results, we find that both type I and type II are broadly consistent with normal galaxies, suggesting no intrinsic difference in the $M_{\bullet}$-$M_{\rm h}$ relation between non-AGN galaxies and AGNs.  

\subsection{Constraint on black-hole mass in simulations}

In state-of-the-art cosmological hydro-dynamical simulations, black-hole growth fed by gas accretion is a crucial factor in driving AGN feedback, which, in turn, is a major mechanism to suppress star formation activities in massive galaxies \citep{SIMBA}. However, these simulations cannot resolve the detailed accretion process. Instead, empirical subgrid recipes are employed to model this process, and the free parameters in these recipes need to be calibrated using observational scaling relations, such as the $M_{\bullet}$-$M_*$ relation \citep{luminosity}.

However, we note that stellar mass itself is subject to several subgrid processes, including the stellar feedback and the AGN feedback, which makes the calibration process quite complicated. In contrast, halo masses are relatively robust and less sensitive to baryonic processes. Therefore, the black-hole-to-halo mass relation is a better scaling relation to calibration subgrid parameters in these simulations, and our work takes the first step to establish this relation in observation.

To compare our measurements with simulations,
we used the RefL0100N1054 run for EAGLE \citep{EAGLE1}, the TNG100 run for IllustrisTNG \citep{TNG}, and the m100n1024 run for SIMBA \citep{SIMBA}.
We calculated AGN luminosity from Eq.~1 in \cite{luminosity} and selected central subhalos with $L_{\rm BH}/L_{\rm Edd}>0.001$ as the ``AGN'' sample in the simulation. We used the snapshot with $z\sim 0.4$, which is the average redshift of our type I AGN sample. We find no significant evolution between $z=0.4$ and $z=0.1$ in the three simulations, which is consistent with our observation. We also compared the ``AGN'' sample and central galaxy sample in the simulation and found no statistically significant difference between their $M_{\bullet}$-$M_{\rm h}$ relations. The AGN $M_{\bullet}$-$M_{\rm h}$ relations from the three simulations are plotted in the right panel of Fig.~\ref{fig:main_res}.  

Although the three simulations calibrate their models to be in good agreement with observed relations \citep{2013ApJ...764..184M,Kormendy&Ho} between $M_{\bullet}$ and stellar mass of the galaxy, $M_*$, or of the bulge, $M_{\rm bulge}$, their $M_{\bullet}$-$M_{\rm h}$ relations do not perfectly match our measurements. The difference among the simulations under similar calibration clearly reflects how different black-hole accretion and AGN feedback mechanisms shape the black-hole masses in simulations. 
The predicted halo mass from EAGLE is consistent with ours at low masses but is significantly higher than ours at $\log M_{\bullet}/M_{\odot}>8$. However, TNG and SIMBA predict lower halo masses at fixed black-hole masses compared to EAGLE, which are more consistent with our observations (both type I and type II).
Among the three simulations, SIMBA has the $M_{\bullet}$-$M_{\rm h}$ relation that is the closest to ours, with all differences within one sigma.

\subsection{Future prospects}

Current data suffers from a small AGN sample size and limited accuracy of black-hole mass estimation. 
Large spectroscopic surveys such as DESI \citep{2013arXiv1308.0847L} and PFS\footnote{https://pfs.ipmu.jp/} will provide larger quasar samples with reliable virial black-hole mass measurements. 
Already now integral field spectroscopy and reverberation mapping observations are improving the virial black-hole mass measurement accuracy. 
From the perspective of weak-lensing data, we will soon get the $4,800$ square degrees shape catalog with photo-$z$s from the completed UNIONS survey. 

Future weak-lensing surveys such as Euclid \citep{2020A&A...642A.191E}, Rubin-LSST \citep{Ivezic_2019}, Roman \citep{2015arXiv150303757S}, CSST \citep{2019ApJ...883..203G}, and WFST \citep{2023arXiv230607590W}, will provide galaxy samples with accurate shape measurements to higher redshifts, covering larger sky areas. This will enable us to measure the $M_{\bullet}$-$M_{\rm h}$ relation with higher accuracy, as well as its dependency on the host galaxy properties.

\begin{acknowledgements}                                           
We thank Eric Jullo, Shiming Gu, Zheng Zheng, and Chris Miller for their helpful discussion.

The authors thank the support from the National Natural Science Foundation of China (NO. 12192224, 11833005, 11890693), the National Key R\&D Program of China (2021YFC2203100), the 111 project for "Observational and Theoretical Research on Dark Matter and Dark Energy" (B23042), CAS Project for Young Scientists in Basic Research, Grant No. YSBR-062.
H. Hildebrandt is supported by a DFG Heisenberg grant (Hi 1495/5-1), the DFG Collaborative Research Center SFB1491, as well as an ERC Consolidator Grant (No. 770935). Anna Wittje is also supported by SFB1491.

We are honored and grateful for the opportunity of observing the Universe from Maunakea and Haleakala, which both have cultural, historical and natural significance in Hawai'i. This work is based on data obtained as part of the Canada-France Imaging Survey, a CFHT large program of the National Research Council of Canada and the French Centre National de la Recherche Scientifique. Based on observations obtained with MegaPrime/MegaCam, a joint project of CFHT and CEA Saclay, at the Canada-France-Hawaii Telescope (CFHT) which is operated by the National Research Council (NRC) of Canada, the Institut National des Science de l’Univers (INSU) of the Centre National de la Recherche Scientifique (CNRS) of France, and the University of Hawaii. This research used the facilities of the Canadian Astronomy Data Centre operated by the National Research Council of Canada with the support of the Canadian Space Agency. This research is based in part on data collected at Subaru Telescope, which is operated by the National Astronomical Observatory of Japan.
Pan-STARRS is a project of the Institute for Astronomy of the University of Hawai'i, and is supported by the NASA SSO Near Earth Observation Program under grants 80NSSC18K0971, NNX14AM74G, NNX12AR65G, NNX13AQ47G, NNX08AR22G, 80NSSC21K1572 and by the State of Hawai'i.

This work was made possible by utilizing the CANDIDE cluster at the Institut d’Astrophysique de Paris, which was funded through grants from the PNCG, CNES, DIM-ACAV, and the Cosmic Dawn Center and maintained by S.~Rouberol.

This work was supported in part by the Canadian Advanced Network for Astronomical Research (CANFAR) and Compute Canada facilities.

Funding for the Sloan Digital Sky Survey has been provided by the Alfred P.~Sloan Foundation, the Heising-Simons Foundation, the National Science Foundation, and the Participating Institutions. SDSS acknowledges support and resources from the Center for High-Performance Computing at the University of Utah. The SDSS web site is \url{www.sdss.org}.

\end{acknowledgements}

\bibliography{BHHMR}

\begin{thebibliography}{}
\expandafter\ifx\csname natexlab\endcsname\relax\def\natexlab#1{#1}\fi
\providecommand{\url}[1]{\href{#1}{#1}}
\providecommand{\dodoi}[1]{doi:~\href{http://doi.org/#1}{\nolinkurl{#1}}}
\providecommand{\doeprint}[1]{\href{http://ascl.net/#1}{\nolinkurl{http://ascl.net/#1}}}
\providecommand{\doarXiv}[1]{\href{https://arxiv.org/abs/#1}{\nolinkurl{https://arxiv.org/abs/#1}}}

\bibitem[{{Baes} {et~al.}(2003){Baes}, {Buyle}, {Hau}, \&
  {Dejonghe}}]{Baes2003}
{Baes}, M., {Buyle}, P., {Hau}, G. K.~T., \& {Dejonghe}, H. 2003, \mnras, 341,
  L44, \dodoi{10.1046/j.1365-8711.2003.06680.x}

\bibitem[{{Baldwin} {et~al.}(1981){Baldwin}, {Phillips}, \&
  {Terlevich}}]{1981PASP...93....5B}
{Baldwin}, J.~A., {Phillips}, M.~M., \& {Terlevich}, R. 1981, \pasp, 93, 5,
  \dodoi{10.1086/130766}

\bibitem[{{Bandara} {et~al.}(2009){Bandara}, {Crampton}, \&
  {Simard}}]{Bandara2009}
{Bandara}, K., {Crampton}, D., \& {Simard}, L. 2009, \apj, 704, 1135,
  \dodoi{10.1088/0004-637X/704/2/1135}

\bibitem[{{Bentz} {et~al.}(2006){Bentz}, {Peterson}, {Pogge}, {Vestergaard}, \&
  {Onken}}]{2006ApJ...644..133B}
{Bentz}, M.~C., {Peterson}, B.~M., {Pogge}, R.~W., {Vestergaard}, M., \&
  {Onken}, C.~A. 2006, \apj, 644, 133, \dodoi{10.1086/503537}

\bibitem[{{Bertin}(2011)}]{2011ASPC..442..435B}
{Bertin}, E. 2011, in Astronomical Society of the Pacific Conference Series,
  Vol. 442, Astronomical Data Analysis Software and Systems XX, ed. I.~N.
  {Evans}, A.~{Accomazzi}, D.~J. {Mink}, \& A.~H. {Rots}, 435

\bibitem[{{Blanton} {et~al.}(2005){Blanton}, {Schlegel}, {Strauss},
  {Brinkmann}, {Finkbeiner}, {Fukugita}, {Gunn}, {Hogg}, {Ivezi{\'c}}, {Knapp},
  {Lupton}, {Munn}, {Schneider}, {Tegmark}, \& {Zehavi}}]{2005AJ....129.2562B}
{Blanton}, M.~R., {Schlegel}, D.~J., {Strauss}, M.~A., {et~al.} 2005, \aj, 129,
  2562, \dodoi{10.1086/429803}

\bibitem[{{Brinchmann} {et~al.}(2004){Brinchmann}, {Charlot}, {White},
  {Tremonti}, {Kauffmann}, {Heckman}, \& {Brinkmann}}]{2004MNRAS.351.1151B}
{Brinchmann}, J., {Charlot}, S., {White}, S.~D.~M., {et~al.} 2004, \mnras, 351,
  1151, \dodoi{10.1111/j.1365-2966.2004.07881.x}

\bibitem[{{Cappellari} {et~al.}(2006){Cappellari}, {Bacon}, {Bureau}, {Damen},
  {Davies}, {de Zeeuw}, {Emsellem}, {Falc{\'o}n-Barroso}, {Krajnovi{\'c}},
  {Kuntschner}, {McDermid}, {Peletier}, {Sarzi}, {van den Bosch}, \& {van de
  Ven}}]{2006MNRAS.366.1126C}
{Cappellari}, M., {Bacon}, R., {Bureau}, M., {et~al.} 2006, \mnras, 366, 1126,
  \dodoi{10.1111/j.1365-2966.2005.09981.x}

\bibitem[{{Dav{\'e}} {et~al.}(2019){Dav{\'e}}, {Angl{\'e}s-Alc{\'a}zar},
  {Narayanan}, {Li}, {Rafieferantsoa}, \& {Appleby}}]{SIMBA}
{Dav{\'e}}, R., {Angl{\'e}s-Alc{\'a}zar}, D., {Narayanan}, D., {et~al.} 2019,
  \mnras, 486, 2827, \dodoi{10.1093/mnras/stz937}

\bibitem[{{Davis} {et~al.}(2019){Davis}, {Graham}, \& {Combes}}]{DGC2019}
{Davis}, B.~L., {Graham}, A.~W., \& {Combes}, F. 2019, \apj, 877, 64,
  \dodoi{10.3847/1538-4357/ab1aa4}

\bibitem[{{Dey} {et~al.}(2019){Dey}, {Schlegel}, {Lang}, {Blum}, {Burleigh},
  {Fan}, {Findlay}, {Finkbeiner}, {Herrera}, {Juneau}, {Landriau}, {Levi},
  {McGreer}, {Meisner}, {Myers}, {Moustakas}, {Nugent}, {Patej}, {Schlafly},
  {Walker}, {Valdes}, {Weaver}, {Y{\`e}che}, {Zou}, {Zhou}, {Abareshi},
  {Abbott}, {Abolfathi}, {Aguilera}, {Alam}, {Allen}, {Alvarez}, {Annis},
  {Ansarinejad}, {Aubert}, {Beechert}, {Bell}, {BenZvi}, {Beutler}, {Bielby},
  {Bolton}, {Brice{\~n}o}, {Buckley-Geer}, {Butler}, {Calamida}, {Carlberg},
  {Carter}, {Casas}, {Castander}, {Choi}, {Comparat}, {Cukanovaite}, {Delubac},
  {DeVries}, {Dey}, {Dhungana}, {Dickinson}, {Ding}, {Donaldson}, {Duan},
  {Duckworth}, {Eftekharzadeh}, {Eisenstein}, {Etourneau}, {Fagrelius},
  {Farihi}, {Fitzpatrick}, {Font-Ribera}, {Fulmer}, {G{\"a}nsicke},
  {Gaztanaga}, {George}, {Gerdes}, {Gontcho}, {Gorgoni}, {Green}, {Guy},
  {Harmer}, {Hernandez}, {Honscheid}, {Huang}, {James}, {Jannuzi}, {Jiang},
  {Joyce}, {Karcher}, {Karkar}, {Kehoe}, {Kneib}, {Kueter-Young}, {Lan},
  {Lauer}, {Le Guillou}, {Le Van Suu}, {Lee}, {Lesser}, {Perreault Levasseur},
  {Li}, {Mann}, {Marshall}, {Mart{\'\i}nez-V{\'a}zquez}, {Martini}, {du Mas des
  Bourboux}, {McManus}, {Meier}, {M{\'e}nard}, {Metcalfe},
  {Mu{\~n}oz-Guti{\'e}rrez}, {Najita}, {Napier}, {Narayan}, {Newman}, {Nie},
  {Nord}, {Norman}, {Olsen}, {Paat}, {Palanque-Delabrouille}, {Peng},
  {Poppett}, {Poremba}, {Prakash}, {Rabinowitz}, {Raichoor}, {Rezaie},
  {Robertson}, {Roe}, {Ross}, {Ross}, {Rudnick}, {Safonova}, {Saha},
  {S{\'a}nchez}, {Savary}, {Schweiker}, {Scott}, {Seo}, {Shan}, {Silva},
  {Slepian}, {Soto}, {Sprayberry}, {Staten}, {Stillman}, {Stupak}, {Summers},
  {Sien Tie}, {Tirado}, {Vargas-Maga{\~n}a}, {Vivas}, {Wechsler}, {Williams},
  {Yang}, {Yang}, {Yapici}, {Zaritsky}, {Zenteno}, {Zhang}, {Zhang}, {Zhou}, \&
  {Zhou}}]{2019AJ....157..168D}
{Dey}, A., {Schlegel}, D.~J., {Lang}, D., {et~al.} 2019, \aj, 157, 168,
  \dodoi{10.3847/1538-3881/ab089d}

\bibitem[{Dressler \& Richstone(1988)}]{dresslerStellarDynamicsNuclei1988}
Dressler, A., \& Richstone, D.~O. 1988, The Astrophysical Journal, 324, 701,
  \dodoi{10.1086/165930}

\bibitem[{{Erben} {et~al.}(2013){Erben}, {Hildebrandt}, {Miller}, {van
  Waerbeke}, {Heymans}, {Hoekstra}, {Kitching}, {Mellier}, {Benjamin}, {Blake},
  {Bonnett}, {Cordes}, {Coupon}, {Fu}, {Gavazzi}, {Gillis}, {Grocutt}, {Gwyn},
  {Holhjem}, {Hudson}, {Kilbinger}, {Kuijken}, {Milkeraitis}, {Rowe},
  {Schrabback}, {Semboloni}, {Simon}, {Smit}, {Toader}, {Vafaei}, {van Uitert},
  \& {Velander}}]{2013MNRAS.433.2545E}
{Erben}, T., {Hildebrandt}, H., {Miller}, L., {et~al.} 2013, \mnras, 433, 2545,
  \dodoi{10.1093/mnras/stt928}

\bibitem[{{Euclid Collaboration} {et~al.}(2020){Euclid Collaboration},
  {Blanchard}, {Camera}, {Carbone}, {Cardone}, {Casas}, {Clesse}, {Ili{\'c}},
  {Kilbinger}, {Kitching}, {Kunz}, {Lacasa}, {Linder}, {Majerotto},
  {Markovi{\v{c}}}, {Martinelli}, {Pettorino}, {Pourtsidou}, {Sakr},
  {S{\'a}nchez}, {Sapone}, {Tutusaus}, {Yahia-Cherif}, {Yankelevich},
  {Andreon}, {Aussel}, {Balaguera-Antol{\'\i}nez}, {Baldi}, {Bardelli},
  {Bender}, {Biviano}, {Bonino}, {Boucaud}, {Bozzo}, {Branchini}, {Brau-Nogue},
  {Brescia}, {Brinchmann}, {Burigana}, {Cabanac}, {Capobianco}, {Cappi},
  {Carretero}, {Carvalho}, {Casas}, {Castander}, {Castellano}, {Cavuoti},
  {Cimatti}, {Cledassou}, {Colodro-Conde}, {Congedo}, {Conselice}, {Conversi},
  {Copin}, {Corcione}, {Coupon}, {Courtois}, {Cropper}, {Da Silva}, {de la
  Torre}, {Di Ferdinando}, {Dubath}, {Ducret}, {Duncan}, {Dupac}, {Dusini},
  {Fabbian}, {Fabricius}, {Farrens}, {Fosalba}, {Fotopoulou}, {Fourmanoit},
  {Frailis}, {Franceschi}, {Franzetti}, {Fumana}, {Galeotta}, {Gillard},
  {Gillis}, {Giocoli}, {G{\'o}mez-Alvarez}, {Graci{\'a}-Carpio}, {Grupp},
  {Guzzo}, {Hoekstra}, {Hormuth}, {Israel}, {Jahnke}, {Keihanen}, {Kermiche},
  {Kirkpatrick}, {Kohley}, {Kubik}, {Kurki-Suonio}, {Ligori}, {Lilje}, {Lloro},
  {Maino}, {Maiorano}, {Marggraf}, {Martinet}, {Marulli}, {Massey},
  {Medinaceli}, {Mei}, {Mellier}, {Metcalf}, {Metge}, {Meylan}, {Moresco},
  {Moscardini}, {Munari}, {Nichol}, {Niemi}, {Nucita}, {Padilla}, {Paltani},
  {Pasian}, {Percival}, {Pires}, {Polenta}, {Poncet}, {Pozzetti}, {Racca},
  {Raison}, {Renzi}, {Rhodes}, {Romelli}, {Roncarelli}, {Rossetti}, {Saglia},
  {Schneider}, {Scottez}, {Secroun}, {Sirri}, {Stanco}, {Starck}, {Sureau},
  {Tallada-Cresp{\'\i}}, {Tavagnacco}, {Taylor}, {Tenti}, {Tereno},
  {Toledo-Moreo}, {Torradeflot}, {Valenziano}, {Vassallo}, {Verdoes Kleijn},
  {Viel}, {Wang}, {Zacchei}, {Zoubian}, \& {Zucca}}]{2020A&A...642A.191E}
{Euclid Collaboration}, {Blanchard}, A., {Camera}, S., {et~al.} 2020, \aap,
  642, A191, \dodoi{10.1051/0004-6361/202038071}

\bibitem[{{Farrens} {et~al.}(2022){Farrens}, {Guinot}, {Kilbinger}, {Liaudat},
  {Baumont}, {Jimenez}, {Peel}, {Pujol}, {Schmitz}, {Starck}, \&
  {Vitorelli}}]{2022A&A...664A.141F}
{Farrens}, S., {Guinot}, A., {Kilbinger}, M., {et~al.} 2022, \aap, 664, A141,
  \dodoi{10.1051/0004-6361/202243970}

\bibitem[{{Ferrarese}(2002)}]{Ferrarese2002}
{Ferrarese}, L. 2002, \apj, 578, 90, \dodoi{10.1086/342308}

\bibitem[{{Ferrarese} \& {Merritt}(2000)}]{Ferrarese2000}
{Ferrarese}, L., \& {Merritt}, D. 2000, \apjl, 539, L9, \dodoi{10.1086/312838}

\bibitem[{{Gao} \& {White}(2007)}]{2007MNRAS.377L...5G}
{Gao}, L., \& {White}, S. D.~M. 2007, \mnras, 377, L5,
  \dodoi{10.1111/j.1745-3933.2007.00292.x}

\bibitem[{{Gebhardt} {et~al.}(2000){Gebhardt}, {Bender}, {Bower}, {Dressler},
  {Faber}, {Filippenko}, {Green}, {Grillmair}, {Ho}, {Kormendy}, {Lauer},
  {Magorrian}, {Pinkney}, {Richstone}, \& {Tremaine}}]{Gebhardt2000}
{Gebhardt}, K., {Bender}, R., {Bower}, G., {et~al.} 2000, \apjl, 539, L13,
  \dodoi{10.1086/312840}

\bibitem[{{Gong} {et~al.}(2019){Gong}, {Liu}, {Cao}, {Chen}, {Fan}, {Li}, {Li},
  {Li}, {Zhang}, \& {Zhan}}]{2019ApJ...883..203G}
{Gong}, Y., {Liu}, X., {Cao}, Y., {et~al.} 2019, \apj, 883, 203,
  \dodoi{10.3847/1538-4357/ab391e}

\bibitem[{{GRAVITY Collaboration} {et~al.}(2024){GRAVITY Collaboration},
  {Amorim}, {Bourdarot}, {Brandner}, {Cao}, {Cl{\'e}net}, {Davies}, {de Zeeuw},
  {Dexter}, {Drescher}, {Eckart}, {Eisenhauer}, {Fabricius}, {Feuchtgruber},
  {F{\"o}rster Schreiber}, {Garcia}, {Genzel}, {Gillessen}, {Gratadour},
  {H{\"o}nig}, {Kishimoto}, {Lacour}, {Lutz}, {Millour}, {Netzer}, {Ott},
  {Paumard}, {Perraut}, {Perrin}, {Peterson}, {Petrucci}, {Pfuhl}, {Prieto},
  {Rabien}, {Rouan}, {Santos}, {Shangguan}, {Shimizu}, {Sternberg},
  {Straubmeier}, {Sturm}, {Tacconi}, {Tristram}, {Widmann}, \&
  {Woillez}}]{2024arXiv240107676G}
{GRAVITY Collaboration}, {Amorim}, A., {Bourdarot}, G., {et~al.} 2024, arXiv
  e-prints, arXiv:2401.07676, \dodoi{10.48550/arXiv.2401.07676}

\bibitem[{{Guinot} {et~al.}(2022){Guinot}, {Kilbinger}, {Farrens}, {Peel},
  {Pujol}, {Schmitz}, {Starck}, {Erben}, {Gavazzi}, {Gwyn}, {Hudson},
  {Hildebrandt}, {Tobias}, {Miller}, {Spitzer}, {Van Waerbeke}, {Cuillandre},
  {Fabbro}, {McConnachie}, \& {Mellier}}]{2022A&A...666A.162G}
{Guinot}, A., {Kilbinger}, M., {Farrens}, S., {et~al.} 2022, \aap, 666, A162,
  \dodoi{10.1051/0004-6361/202141847}

\bibitem[{{Guzik} \& {Seljak}(2002)}]{GS02}
{Guzik}, J., \& {Seljak}, U. 2002, \mnras, 335, 311,
  \dodoi{10.1046/j.1365-8711.2002.05591.x}

\bibitem[{{Habouzit} {et~al.}(2021){Habouzit}, {Li}, {Somerville}, {Genel},
  {Pillepich}, {Volonteri}, {Dav{\'e}}, {Rosas-Guevara}, {McAlpine}, {Peirani},
  {Hernquist}, {Angl{\'e}s-Alc{\'a}zar}, {Reines}, {Bower}, {Dubois}, {Nelson},
  {Pichon}, \& {Vogelsberger}}]{luminosity}
{Habouzit}, M., {Li}, Y., {Somerville}, R.~S., {et~al.} 2021, \mnras, 503,
  1940, \dodoi{10.1093/mnras/stab496}

\bibitem[{{Heymans} {et~al.}(2012){Heymans}, {Van Waerbeke}, {Miller}, {Erben},
  {Hildebrandt}, {Hoekstra}, {Kitching}, {Mellier}, {Simon}, {Bonnett},
  {Coupon}, {Fu}, {Harnois D{\'e}raps}, {Hudson}, {Kilbinger}, {Kuijken},
  {Rowe}, {Schrabback}, {Semboloni}, {van Uitert}, {Vafaei}, \&
  {Velander}}]{2012MNRAS.427..146H}
{Heymans}, C., {Van Waerbeke}, L., {Miller}, L., {et~al.} 2012, \mnras, 427,
  146, \dodoi{10.1111/j.1365-2966.2012.21952.x}

\bibitem[{{Hildebrandt} {et~al.}(2012){Hildebrandt}, {Erben}, {Kuijken}, {van
  Waerbeke}, {Heymans}, {Coupon}, {Benjamin}, {Bonnett}, {Fu}, {Hoekstra},
  {Kitching}, {Mellier}, {Miller}, {Velander}, {Hudson}, {Rowe}, {Schrabback},
  {Semboloni}, \& {Ben{\'\i}tez}}]{2012MNRAS.421.2355H}
{Hildebrandt}, H., {Erben}, T., {Kuijken}, K., {et~al.} 2012, \mnras, 421,
  2355, \dodoi{10.1111/j.1365-2966.2012.20468.x}

\bibitem[{{Hirata} {et~al.}(2004){Hirata}, {Mandelbaum}, {Seljak}, {Guzik},
  {Padmanabhan}, {Blake}, {Brinkmann}, {Bud{\'a}vari}, {Connolly}, {Csabai},
  {Scranton}, \& {Szalay}}]{2004MNRAS.353..529H}
{Hirata}, C.~M., {Mandelbaum}, R., {Seljak}, U., {et~al.} 2004, \mnras, 353,
  529, \dodoi{10.1111/j.1365-2966.2004.08090.x}

\bibitem[{{Kauffmann} {et~al.}(2003){Kauffmann}, {Heckman}, {White}, {Charlot},
  {Tremonti}, {Brinchmann}, {Bruzual}, {Peng}, {Seibert}, {Bernardi},
  {Blanton}, {Brinkmann}, {Castander}, {Cs{\'a}bai}, {Fukugita}, {Ivezic},
  {Munn}, {Nichol}, {Padmanabhan}, {Thakar}, {Weinberg}, \&
  {York}}]{2003MNRAS.341...33K}
{Kauffmann}, G., {Heckman}, T.~M., {White}, S. D.~M., {et~al.} 2003, \mnras,
  341, 33, \dodoi{10.1046/j.1365-8711.2003.06291.x}

\bibitem[{{Kilbinger}(2015)}]{K15}
{Kilbinger}, M. 2015, Reports on Progress in Physics, 78, 086901,
  \dodoi{10.1088/0034-4885/78/8/086901}

\bibitem[{Kohonen(1982)}]{kohonen1982self}
Kohonen, T. 1982, Biological cybernetics, 43, 59

\bibitem[{{Kormendy} \& {Bender}(2011)}]{K&B2011}
{Kormendy}, J., \& {Bender}, R. 2011, \nat, 469, 377,
  \dodoi{10.1038/nature09695}

\bibitem[{Kormendy \& Ho(2013)}]{kormendyCoevolutionNotSupermassive2013}
Kormendy, J., \& Ho, L.~C. 2013, Annual Review of Astronomy and Astrophysics,
  51, 511, \dodoi{10.1146/annurev-astro-082708-101811}

\bibitem[{{Kormendy} \& {Ho}(2013)}]{Kormendy&Ho}
{Kormendy}, J., \& {Ho}, L.~C. 2013, \araa, 51, 511,
  \dodoi{10.1146/annurev-astro-082708-101811}

\bibitem[{{Kormendy} \& {Richstone}(1995)}]{Kormendy1995}
{Kormendy}, J., \& {Richstone}, D. 1995, \araa, 33, 581,
  \dodoi{10.1146/annurev.aa.33.090195.003053}

\bibitem[{{Krumpe} {et~al.}(2015){Krumpe}, {Miyaji}, {Husemann}, {Fanidakis},
  {Coil}, \& {Aceves}}]{Krumpe2015}
{Krumpe}, M., {Miyaji}, T., {Husemann}, B., {et~al.} 2015, \apj, 815, 21,
  \dodoi{10.1088/0004-637X/815/1/21}

\bibitem[{Krumpe {et~al.}(2023)Krumpe, Miyaji, Georgakakis, Schulze, Coil,
  Dwelly, Coffey, Comparat, Aceves, Salvato, Merloni, Maraston, Nandra,
  Brownstein, \& Schneider}]{krumpe2023spatial}
Krumpe, M., Miyaji, T., Georgakakis, A., {et~al.} 2023, The spatial clustering
  of ROSAT all-sky survey Active Galactic Nuclei: V. The evolution of
  broad-line AGN clustering properties in the last 6 Gyr.
\newblock \doarXiv{2304.02036}

\bibitem[{{Le F{\`e}vre} {et~al.}(2005){Le F{\`e}vre}, {Vettolani}, {Garilli},
  {Tresse}, {Bottini}, {Le Brun}, {Maccagni}, {Picat}, {Scaramella},
  {Scodeggio}, {Zanichelli}, {Adami}, {Arnaboldi}, {Arnouts}, {Bardelli},
  {Bolzonella}, {Cappi}, {Charlot}, {Ciliegi}, {Contini}, {Foucaud},
  {Franzetti}, {Gavignaud}, {Guzzo}, {Ilbert}, {Iovino}, {McCracken}, {Marano},
  {Marinoni}, {Mathez}, {Mazure}, {Meneux}, {Merighi}, {Paltani}, {Pell{\`o}},
  {Pollo}, {Pozzetti}, {Radovich}, {Zamorani}, {Zucca}, {Bondi}, {Bongiorno},
  {Busarello}, {Lamareille}, {Mellier}, {Merluzzi}, {Ripepi}, \&
  {Rizzo}}]{2005A&A...439..845L}
{Le F{\`e}vre}, O., {Vettolani}, G., {Garilli}, B., {et~al.} 2005, \aap, 439,
  845, \dodoi{10.1051/0004-6361:20041960}

\bibitem[{{Leauthaud} {et~al.}(2015){Leauthaud}, {J. Benson}, {Civano}, {L.
  Coil}, {Bundy}, {Massey}, {Schramm}, {Schulze}, {Capak}, {Elvis}, {Kulier},
  \& {Rhodes}}]{2015MNRAS.446.1874L}
{Leauthaud}, A., {J. Benson}, A., {Civano}, F., {et~al.} 2015, \mnras, 446,
  1874, \dodoi{10.1093/mnras/stu2210}

\bibitem[{{Levi} {et~al.}(2013){Levi}, {Bebek}, {Beers}, {Blum}, {Cahn},
  {Eisenstein}, {Flaugher}, {Honscheid}, {Kron}, {Lahav}, {McDonald}, {Roe},
  {Schlegel}, \& {representing the DESI collaboration}}]{2013arXiv1308.0847L}
{Levi}, M., {Bebek}, C., {Beers}, T., {et~al.} 2013, arXiv e-prints,
  arXiv:1308.0847, \dodoi{10.48550/arXiv.1308.0847}

\bibitem[{{Liaudat} {et~al.}(2021){Liaudat}, {Bonnin}, {Starck}, {Schmitz},
  {Guinot}, {Kilbinger}, \& {Gwyn}}]{MCCD21}
{Liaudat}, T., {Bonnin}, J., {Starck}, J.-L., {et~al.} 2021, \aap, 646, A27,
  \dodoi{10.1051/0004-6361/202039584}

\bibitem[{{Liu} {et~al.}(2019){Liu}, {Liu}, {Dong}, {Zhou}, {Wang}, {Lu}, \&
  {Yuan}}]{Liu2019}
{Liu}, H.-Y., {Liu}, W.-J., {Dong}, X.-B., {et~al.} 2019, \apjs, 243, 21,
  \dodoi{10.3847/1538-4365/ab298b}

\bibitem[{{Luo} {et~al.}(2018){Luo}, {Yang}, {Lu}, {Shi}, {Zhang}, {Mo}, {Shu},
  {Fu}, {Radovich}, {Zhang}, {Li}, {Sunayama}, \& {Wang}}]{2018ApJ...862....4L}
{Luo}, W., {Yang}, X., {Lu}, T., {et~al.} 2018, \apj, 862, 4,
  \dodoi{10.3847/1538-4357/aacaf1}

\bibitem[{{Luo} {et~al.}(2022){Luo}, {Silverman}, {More}, {Goulding},
  {Miyatake}, {Nishimichi}, {Hikage}, {Kawinwanichakij}, {Li}, {Li},
  {Medezinski}, {Oguri}, {Oogi}, \& {Sifon}}]{2022arXiv220403817L}
{Luo}, W., {Silverman}, J.~D., {More}, S., {et~al.} 2022, arXiv e-prints,
  arXiv:2204.03817, \dodoi{10.48550/arXiv.2204.03817}

\bibitem[{Lyke {et~al.}(2020)Lyke, Higley, McLane, Schurhammer, Myers, Ross,
  Dawson, Chabanier, Martini, Busca, du~Mas~des Bourboux, Salvato,
  Streblyanska, Zarrouk, Burtin, Anderson, Bautista, Bizyaev, Brandt,
  Brinkmann, Brownstein, Comparat, Green, de~la Macorra, Gutiérrez, Hou,
  Newman, Palanque-Delabrouille, Pâris, Percival, Petitjean, Rich, Rossi,
  Schneider, Smith, Vivek, \& Weaver}]{Lyke2020}
Lyke, B.~W., Higley, A.~N., McLane, J.~N., {et~al.} 2020, The Astrophysical
  Journal Supplement Series, 250, 8, \dodoi{10.3847/1538-4365/aba623}

\bibitem[{{Magorrian} {et~al.}(1998){Magorrian}, {Tremaine}, {Richstone},
  {Bender}, {Bower}, {Dressler}, {Faber}, {Gebhardt}, {Green}, {Grillmair},
  {Kormendy}, \& {Lauer}}]{1998AJ....115.2285M}
{Magorrian}, J., {Tremaine}, S., {Richstone}, D., {et~al.} 1998, \aj, 115,
  2285, \dodoi{10.1086/300353}

\bibitem[{{Mandelbaum} {et~al.}(2009){Mandelbaum}, {Li}, {Kauffmann}, \&
  {White}}]{2009MNRAS.393..377M}
{Mandelbaum}, R., {Li}, C., {Kauffmann}, G., \& {White}, S. D.~M. 2009, \mnras,
  393, 377, \dodoi{10.1111/j.1365-2966.2008.14235.x}

\bibitem[{{Mandelbaum} {et~al.}(2006){Mandelbaum}, {Seljak}, {Kauffmann},
  {Hirata}, \& {Brinkmann}}]{2006MNRAS.368..715M}
{Mandelbaum}, R., {Seljak}, U., {Kauffmann}, G., {Hirata}, C.~M., \&
  {Brinkmann}, J. 2006, \mnras, 368, 715,
  \dodoi{10.1111/j.1365-2966.2006.10156.x}

\bibitem[{{Marasco} {et~al.}(2021){Marasco}, {Cresci}, {Posti}, {Fraternali},
  {Mannucci}, {Marconi}, {Belfiore}, \& {Fall}}]{Marasco2021}
{Marasco}, A., {Cresci}, G., {Posti}, L., {et~al.} 2021, \mnras, 507, 4274,
  \dodoi{10.1093/mnras/stab2317}

\bibitem[{{Masters} {et~al.}(2015){Masters}, {Capak}, {Stern}, {Ilbert},
  {Salvato}, {Schmidt}, {Longo}, {Rhodes}, {Paltani}, {Mobasher}, {Hoekstra},
  {Hildebrandt}, {Coupon}, {Steinhardt}, {Speagle}, {Faisst}, {Kalinich},
  {Brodwin}, {Brescia}, \& {Cavuoti}}]{2015ApJ...813...53M}
{Masters}, D., {Capak}, P., {Stern}, D., {et~al.} 2015, \apj, 813, 53,
  \dodoi{10.1088/0004-637X/813/1/53}

\bibitem[{{McConnell} \& {Ma}(2013)}]{2013ApJ...764..184M}
{McConnell}, N.~J., \& {Ma}, C.-P. 2013, \apj, 764, 184,
  \dodoi{10.1088/0004-637X/764/2/184}

\bibitem[{{Miller} {et~al.}(2007){Miller}, {Kitching}, {Heymans}, {Heavens}, \&
  {van Waerbeke}}]{2007MNRAS.382..315M}
{Miller}, L., {Kitching}, T.~D., {Heymans}, C., {Heavens}, A.~F., \& {van
  Waerbeke}, L. 2007, \mnras, 382, 315,
  \dodoi{10.1111/j.1365-2966.2007.12363.x}

\bibitem[{{Newman} {et~al.}(2013){Newman}, {Cooper}, {Davis}, {Faber}, {Coil},
  {Guhathakurta}, {Koo}, {Phillips}, {Conroy}, {Dutton}, {Finkbeiner}, {Gerke},
  {Rosario}, {Weiner}, {Willmer}, {Yan}, {Harker}, {Kassin}, {Konidaris},
  {Lai}, {Madgwick}, {Noeske}, {Wirth}, {Connolly}, {Kaiser}, {Kirby},
  {Lemaux}, {Lin}, {Lotz}, {Luppino}, {Marinoni}, {Matthews}, {Metevier}, \&
  {Schiavon}}]{2013ApJS..208....5N}
{Newman}, J.~A., {Cooper}, M.~C., {Davis}, M., {et~al.} 2013, \apjs, 208, 5,
  \dodoi{10.1088/0067-0049/208/1/5}

\bibitem[{{Pizzella} {et~al.}(2005){Pizzella}, {Corsini}, {Dalla Bont{\`a}},
  {Sarzi}, {Coccato}, \& {Bertola}}]{Pizzella2005}
{Pizzella}, A., {Corsini}, E.~M., {Dalla Bont{\`a}}, E., {et~al.} 2005, \apj,
  631, 785, \dodoi{10.1086/430513}

\bibitem[{{Planck Collaboration} {et~al.}(2020){Planck Collaboration},
  {Aghanim}, {Akrami}, {Ashdown}, {Aumont}, {Baccigalupi}, {Ballardini},
  {Banday}, {Barreiro}, {Bartolo}, {Basak}, {Battye}, {Benabed}, {Bernard},
  {Bersanelli}, {Bielewicz}, {Bock}, {Bond}, {Borrill}, {Bouchet}, {Boulanger},
  {Bucher}, {Burigana}, {Butler}, {Calabrese}, {Cardoso}, {Carron},
  {Challinor}, {Chiang}, {Chluba}, {Colombo}, {Combet}, {Contreras}, {Crill},
  {Cuttaia}, {de Bernardis}, {de Zotti}, {Delabrouille}, {Delouis}, {Di
  Valentino}, {Diego}, {Dor{\'e}}, {Douspis}, {Ducout}, {Dupac}, {Dusini},
  {Efstathiou}, {Elsner}, {En{\ss}lin}, {Eriksen}, {Fantaye}, {Farhang},
  {Fergusson}, {Fernandez-Cobos}, {Finelli}, {Forastieri}, {Frailis},
  {Fraisse}, {Franceschi}, {Frolov}, {Galeotta}, {Galli}, {Ganga},
  {G{\'e}nova-Santos}, {Gerbino}, {Ghosh}, {Gonz{\'a}lez-Nuevo}, {G{\'o}rski},
  {Gratton}, {Gruppuso}, {Gudmundsson}, {Hamann}, {Handley}, {Hansen},
  {Herranz}, {Hildebrandt}, {Hivon}, {Huang}, {Jaffe}, {Jones}, {Karakci},
  {Keih{\"a}nen}, {Keskitalo}, {Kiiveri}, {Kim}, {Kisner}, {Knox},
  {Krachmalnicoff}, {Kunz}, {Kurki-Suonio}, {Lagache}, {Lamarre}, {Lasenby},
  {Lattanzi}, {Lawrence}, {Le Jeune}, {Lemos}, {Lesgourgues}, {Levrier},
  {Lewis}, {Liguori}, {Lilje}, {Lilley}, {Lindholm}, {L{\'o}pez-Caniego},
  {Lubin}, {Ma}, {Mac{\'\i}as-P{\'e}rez}, {Maggio}, {Maino}, {Mandolesi},
  {Mangilli}, {Marcos-Caballero}, {Maris}, {Martin}, {Martinelli},
  {Mart{\'\i}nez-Gonz{\'a}lez}, {Matarrese}, {Mauri}, {McEwen}, {Meinhold},
  {Melchiorri}, {Mennella}, {Migliaccio}, {Millea}, {Mitra},
  {Miville-Desch{\^e}nes}, {Molinari}, {Montier}, {Morgante}, {Moss}, {Natoli},
  {N{\o}rgaard-Nielsen}, {Pagano}, {Paoletti}, {Partridge}, {Patanchon},
  {Peiris}, {Perrotta}, {Pettorino}, {Piacentini}, {Polastri}, {Polenta},
  {Puget}, {Rachen}, {Reinecke}, {Remazeilles}, {Renzi}, {Rocha}, {Rosset},
  {Roudier}, {Rubi{\~n}o-Mart{\'\i}n}, {Ruiz-Granados}, {Salvati}, {Sandri},
  {Savelainen}, {Scott}, {Shellard}, {Sirignano}, {Sirri}, {Spencer},
  {Sunyaev}, {Suur-Uski}, {Tauber}, {Tavagnacco}, {Tenti}, {Toffolatti},
  {Tomasi}, {Trombetti}, {Valenziano}, {Valiviita}, {Van Tent}, {Vibert},
  {Vielva}, {Villa}, {Vittorio}, {Wandelt}, {Wehus}, {White}, {White},
  {Zacchei}, \& {Zonca}}]{Planck18}
{Planck Collaboration}, {Aghanim}, N., {Akrami}, Y., {et~al.} 2020, \aap, 641,
  A6, \dodoi{10.1051/0004-6361/201833910}

\bibitem[{{Powell} {et~al.}(2018){Powell}, {Cappelluti}, {Urry}, {Koss},
  {Finoguenov}, {Ricci}, {Trakhtenbrot}, {Allevato}, {Ajello}, {Oh},
  {Schawinski}, \& {Secrest}}]{Powell2018}
{Powell}, M.~C., {Cappelluti}, N., {Urry}, C.~M., {et~al.} 2018, \apj, 858,
  110, \dodoi{10.3847/1538-4357/aabd7f}

\bibitem[{{Powell} {et~al.}(2022){Powell}, {Allen}, {Caglar}, {Cappelluti},
  {Harrison}, {Irving}, {Koss}, {Mantz}, {Oh}, {Ricci}, {Shaper}, {Stern},
  {Trakhtenbrot}, {Urry}, \& {Wong}}]{Powell2022}
{Powell}, M.~C., {Allen}, S.~W., {Caglar}, T., {et~al.} 2022, arXiv e-prints,
  arXiv:2209.02728.
\newblock \doarXiv{2209.02728}

\bibitem[{{Robinson} {et~al.}(2021){Robinson}, {Bentz}, {Courtois}, {Johnson},
  {Crenshaw}, {Meena}, {Polack}, {Silverstein}, \& {Chen}}]{Robinson2021}
{Robinson}, J.~H., {Bentz}, M.~C., {Courtois}, H.~M., {et~al.} 2021, \apj, 912,
  160, \dodoi{10.3847/1538-4357/abedaa}

\bibitem[{Sabra {et~al.}(2015)Sabra, Saliba, Akl, \& Chahine}]{Sabra2015}
Sabra, B.~M., Saliba, C., Akl, M.~A., \& Chahine, G. 2015, The Astrophysical
  Journal, 803, 5, \dodoi{10.1088/0004-637X/803/1/5}

\bibitem[{{Saglia} {et~al.}(2016){Saglia}, {Opitsch}, {Erwin}, {Thomas},
  {Beifiori}, {Fabricius}, {Mazzalay}, {Nowak}, {Rusli}, \&
  {Bender}}]{Saglia2016}
{Saglia}, R.~P., {Opitsch}, M., {Erwin}, P., {et~al.} 2016, \apj, 818, 47,
  \dodoi{10.3847/0004-637X/818/1/47}

\bibitem[{{Schaye} {et~al.}(2015){Schaye}, {Crain}, {Bower}, {Furlong},
  {Schaller}, {Theuns}, {Dalla Vecchia}, {Frenk}, {McCarthy}, {Helly},
  {Jenkins}, {Rosas-Guevara}, {White}, {Baes}, {Booth}, {Camps}, {Navarro},
  {Qu}, {Rahmati}, {Sawala}, {Thomas}, \& {Trayford}}]{EAGLE1}
{Schaye}, J., {Crain}, R.~A., {Bower}, R.~G., {et~al.} 2015, \mnras, 446, 521,
  \dodoi{10.1093/mnras/stu2058}

\bibitem[{{Scodeggio} {et~al.}(2018){Scodeggio}, {Guzzo}, {Garilli}, {Granett},
  {Bolzonella}, {de la Torre}, {Abbas}, {Adami}, {Arnouts}, {Bottini}, {Cappi},
  {Coupon}, {Cucciati}, {Davidzon}, {Franzetti}, {Fritz}, {Iovino}, {Krywult},
  {Le Brun}, {Le F{\`e}vre}, {Maccagni}, {Ma{\l}ek}, {Marchetti}, {Marulli},
  {Polletta}, {Pollo}, {Tasca}, {Tojeiro}, {Vergani}, {Zanichelli}, {Bel},
  {Branchini}, {De Lucia}, {Ilbert}, {McCracken}, {Moutard}, {Peacock},
  {Zamorani}, {Burden}, {Fumana}, {Jullo}, {Marinoni}, {Mellier}, {Moscardini},
  \& {Percival}}]{2018A&A...609A..84S}
{Scodeggio}, M., {Guzzo}, L., {Garilli}, B., {et~al.} 2018, \aap, 609, A84,
  \dodoi{10.1051/0004-6361/201630114}

\bibitem[{{Shankar} {et~al.}(2020){Shankar}, {Allevato}, {Bernardi}, {Marsden},
  {Lapi}, {Menci}, {Grylls}, {Krumpe}, {Zanisi}, {Ricci}, {La Franca}, {Baldi},
  {Moreno}, \& {Sheth}}]{Shankar2020}
{Shankar}, F., {Allevato}, V., {Bernardi}, M., {et~al.} 2020, Nature Astronomy,
  4, 282, \dodoi{10.1038/s41550-019-0949-y}

\bibitem[{{Shen}(2013)}]{Shen2013}
{Shen}, Y. 2013, Bulletin of the Astronomical Society of India, 41, 61,
  \dodoi{10.48550/arXiv.1302.2643}

\bibitem[{{Shen} {et~al.}(2011){Shen}, {Richards}, {Strauss}, {Hall},
  {Schneider}, {Snedden}, {Bizyaev}, {Brewington}, {Malanushenko},
  {Malanushenko}, {Oravetz}, {Pan}, \& {Simmons}}]{Shen11}
{Shen}, Y., {Richards}, G.~T., {Strauss}, M.~A., {et~al.} 2011, \apjs, 194, 45,
  \dodoi{10.1088/0067-0049/194/2/45}

\bibitem[{Shimasaku \& Izumi(2019)}]{Shimasaku2019}
Shimasaku, K., \& Izumi, T. 2019, The Astrophysical Journal Letters, 872, L29,
  \dodoi{10.3847/2041-8213/ab053f}

\bibitem[{{Spergel} {et~al.}(2015){Spergel}, {Gehrels}, {Baltay}, {Bennett},
  {Breckinridge}, {Donahue}, {Dressler}, {Gaudi}, {Greene}, {Guyon}, {Hirata},
  {Kalirai}, {Kasdin}, {Macintosh}, {Moos}, {Perlmutter}, {Postman},
  {Rauscher}, {Rhodes}, {Wang}, {Weinberg}, {Benford}, {Hudson}, {Jeong},
  {Mellier}, {Traub}, {Yamada}, {Capak}, {Colbert}, {Masters}, {Penny},
  {Savransky}, {Stern}, {Zimmerman}, {Barry}, {Bartusek}, {Carpenter}, {Cheng},
  {Content}, {Dekens}, {Demers}, {Grady}, {Jackson}, {Kuan}, {Kruk}, {Melton},
  {Nemati}, {Parvin}, {Poberezhskiy}, {Peddie}, {Ruffa}, {Wallace}, {Whipple},
  {Wollack}, \& {Zhao}}]{2015arXiv150303757S}
{Spergel}, D., {Gehrels}, N., {Baltay}, C., {et~al.} 2015, arXiv e-prints,
  arXiv:1503.03757, \dodoi{10.48550/arXiv.1503.03757}

\bibitem[{Springel {et~al.}(2017)Springel, Pakmor, Pillepich, Weinberger,
  Nelson, Hernquist, Vogelsberger, Genel, Torrey, Marinacci, \& Naiman}]{TNG}
Springel, V., Pakmor, R., Pillepich, A., {et~al.} 2017, Monthly Notices of the
  Royal Astronomical Society, 475, 676, \dodoi{10.1093/mnras/stx3304}

\bibitem[{{Tinker} {et~al.}(2008){Tinker}, {Kravtsov}, {Klypin}, {Abazajian},
  {Warren}, {Yepes}, {Gottl{\"o}ber}, \& {Holz}}]{TinkerHMF}
{Tinker}, J., {Kravtsov}, A.~V., {Klypin}, A., {et~al.} 2008, \apj, 688, 709,
  \dodoi{10.1086/591439}

\bibitem[{{Tinker} {et~al.}(2010){Tinker}, {Robertson}, {Kravtsov}, {Klypin},
  {Warren}, {Yepes}, \& {Gottl{\"o}ber}}]{Tinker2010}
{Tinker}, J.~L., {Robertson}, B.~E., {Kravtsov}, A.~V., {et~al.} 2010, \apj,
  724, 878, \dodoi{10.1088/0004-637X/724/2/878}

\bibitem[{{Vestergaard} \& {Peterson}(2006)}]{2006ApJ...641..689V}
{Vestergaard}, M., \& {Peterson}, B.~M. 2006, \apj, 641, 689,
  \dodoi{10.1086/500572}

\bibitem[{{Volonteri} {et~al.}(2011){Volonteri}, {Natarajan}, \&
  {G{\"u}ltekin}}]{VNG2011}
{Volonteri}, M., {Natarajan}, P., \& {G{\"u}ltekin}, K. 2011, \apj, 737, 50,
  \dodoi{10.1088/0004-637X/737/2/50}

\bibitem[{{Wang} {et~al.}(2024){Wang}, {Mo}, {Chen}, {Wang}, {Yang}, {Wang},
  {Peng}, \& {Cai}}]{2024MNRAS.528.2046W}
{Wang}, K., {Mo}, H.~J., {Chen}, Y., {et~al.} 2024, \mnras, 528, 2046,
  \dodoi{10.1093/mnras/stae163}

\bibitem[{Wechsler \& Tinker(2018)}]{doi:10.1146/annurev-astro-081817-051756}
Wechsler, R.~H., \& Tinker, J.~L. 2018, Annual Review of Astronomy and
  Astrophysics, 56, 435, \dodoi{10.1146/annurev-astro-081817-051756}

\bibitem[{{WFST Collaboration} {et~al.}(2023){WFST Collaboration}, {Wang},
  {Liu}, {Cai}, {Gen}, {Fang}, {He}, {Jiang}, {Jiang}, {Kong}, {Li}, {Li},
  {Luo}, {Pan}, {Wu}, {Yang}, {Yu}, {Zheng}, {Zhu}, {Cai}, {Chen}, {Chen},
  {Dai}, {Fan}, {Fan}, {Fang}, {He}, {Hu}, {Hu}, {Jin}, {Jiang}, {Li}, {Li},
  {Li}, {Liang}, {Lin}, {Liu}, {Liu}, {Liu}, {Liu}, {Liu}, {Lou}, {Qu},
  {Sheng}, {Shi}, {Shu}, {Su}, {Sun}, {Wang}, {Wang}, {Wang}, {Wang}, {Wei},
  {Wei}, {Xue}, {Yan}, {Yang}, {Yuan}, {Yuan}, {Zhang}, {Zhang}, {Zhao}, \&
  {Zhao}}]{2023arXiv230607590W}
{WFST Collaboration}, {Wang}, T., {Liu}, G., {et~al.} 2023, arXiv e-prints,
  arXiv:2306.07590, \dodoi{10.48550/arXiv.2306.07590}

\bibitem[{{Wright} {et~al.}(2020){Wright}, {Hildebrandt}, {van den Busch}, \&
  {Heymans}}]{2020A&A...637A.100W}
{Wright}, A.~H., {Hildebrandt}, H., {van den Busch}, J.~L., \& {Heymans}, C.
  2020, \aap, 637, A100, \dodoi{10.1051/0004-6361/201936782}

\bibitem[{{Wu} \& {Shen}(2022)}]{Wu&Shen2022}
{Wu}, Q., \& {Shen}, Y. 2022, \apjs, 263, 42, \dodoi{10.3847/1538-4365/ac9ead}

\bibitem[{{Yang} {et~al.}(2008){Yang}, {Mo}, \& {van den
  Bosch}}]{2008ApJ...676..248Y}
{Yang}, X., {Mo}, H.~J., \& {van den Bosch}, F.~C. 2008, \apj, 676, 248,
  \dodoi{10.1086/528954}

\bibitem[{{Yang} {et~al.}(2007){Yang}, {Mo}, {van den Bosch}, {Pasquali}, {Li},
  \& {Barden}}]{2007ApJ...671..153Y}
{Yang}, X., {Mo}, H.~J., {van den Bosch}, F.~C., {et~al.} 2007, \apj, 671, 153,
  \dodoi{10.1086/522027}

\bibitem[{{Zhang} {et~al.}(2023{\natexlab{a}}){Zhang}, {Behroozi}, {Volonteri},
  {Silk}, {Fan}, {Aird}, {Yang}, \& {Hopkins}}]{TRINITYIII}
{Zhang}, H., {Behroozi}, P., {Volonteri}, M., {et~al.} 2023{\natexlab{a}},
  arXiv e-prints, arXiv:2305.19315, \dodoi{10.48550/arXiv.2305.19315}

\bibitem[{{Zhang} {et~al.}(2023{\natexlab{b}}){Zhang}, {Behroozi}, {Volonteri},
  {Silk}, {Fan}, {Hopkins}, {Yang}, \& {Aird}}]{TRINITYI}
---. 2023{\natexlab{b}}, \mnras, 518, 2123, \dodoi{10.1093/mnras/stac2633}

\bibitem[{{Zhang} {et~al.}(2022){Zhang}, {Liu}, {Vaquero}, {Li}, {Wang},
  {Shen}, \& {Dong}}]{2022AJ....164..128Z}
{Zhang}, J., {Liu}, C., {Vaquero}, P.~A., {et~al.} 2022, \aj, 164, 128,
  \dodoi{10.3847/1538-3881/ac84d8}

\bibitem[{{Zhang} {et~al.}(2023{\natexlab{c}}){Zhang}, {Wang}, {Luo}, {Mo},
  {Zhang}, {Yang}, {Li}, \& {Li}}]{Zhang2023}
{Zhang}, Z., {Wang}, H., {Luo}, W., {et~al.} 2023{\natexlab{c}}, arXiv
  e-prints, arXiv:2305.06803, \dodoi{10.48550/arXiv.2305.06803}

\bibitem[{Željko Ivezić {et~al.}(2019)Željko Ivezić, Kahn, Tyson, Abel,
  Acosta, Allsman, Alonso, AlSayyad, Anderson, Andrew, Angel, Angeli, Ansari,
  Antilogus, Araujo, Armstrong, Arndt, Astier, Éric Aubourg, Auza, Axelrod,
  Bard, Barr, Barrau, Bartlett, Bauer, Bauman, Baumont, Bechtol, Bechtol,
  Becker, Becla, Beldica, Bellavia, Bianco, Biswas, Blanc, Blazek, Blandford,
  Bloom, Bogart, Bond, Booth, Borgland, Borne, Bosch, Boutigny, Brackett,
  Bradshaw, Brandt, Brown, Bullock, Burchat, Burke, Cagnoli, Calabrese,
  Callahan, Callen, Carlin, Carlson, Chandrasekharan, Charles-Emerson, Chesley,
  Cheu, Chiang, Chiang, Chirino, Chow, Ciardi, Claver, Cohen-Tanugi, Cockrum,
  Coles, Connolly, Cook, Cooray, Covey, Cribbs, Cui, Cutri, Daly, Daniel,
  Daruich, Daubard, Daues, Dawson, Delgado, Dellapenna, de~Peyster,
  de~Val-Borro, Digel, Doherty, Dubois, Dubois-Felsmann, Durech, Economou,
  Eifler, Eracleous, Emmons, Neto, Ferguson, Figueroa, Fisher-Levine, Focke,
  Foss, Frank, Freemon, Gangler, Gawiser, Geary, Gee, Geha, Gessner, Gibson,
  Gilmore, Glanzman, Glick, Goldina, Goldstein, Goodenow, Graham, Gressler,
  Gris, Guy, Guyonnet, Haller, Harris, Hascall, Haupt, Hernandez, Herrmann,
  Hileman, Hoblitt, Hodgson, Hogan, Howard, Huang, Huffer, Ingraham, Innes,
  Jacoby, Jain, Jammes, Jee, Jenness, Jernigan, Jevremović, Johns, Johnson,
  Johnson, Jones, Juramy-Gilles, Jurić, Kalirai, Kallivayalil, Kalmbach,
  Kantor, Karst, Kasliwal, Kelly, Kessler, Kinnison, Kirkby, Knox, Kotov,
  Krabbendam, Krughoff, Kubánek, Kuczewski, Kulkarni, Ku, Kurita, Lage,
  Lambert, Lange, Langton, Guillou, Levine, Liang, Lim, Lintott, Long, Lopez,
  Lotz, Lupton, Lust, MacArthur, Mahabal, Mandelbaum, Markiewicz, Marsh,
  Marshall, Marshall, May, McKercher, McQueen, Meyers, Migliore, Miller, Mills,
  Miraval, Moeyens, Moolekamp, Monet, Moniez, Monkewitz, Montgomery, Morrison,
  Mueller, Muller, Arancibia, Neill, Newbry, Nief, Nomerotski, Nordby,
  O’Connor, Oliver, Olivier, Olsen, O’Mullane, Ortiz, Osier, Owen, Pain,
  Palecek, Parejko, Parsons, Pease, Peterson, Peterson, Petravick, Petrick,
  Petry, Pierfederici, Pietrowicz, Pike, Pinto, Plante, Plate, Plutchak, Price,
  Prouza, Radeka, Rajagopal, Rasmussen, Regnault, Reil, Reiss, Reuter, Ridgway,
  Riot, Ritz, Robinson, Roby, Roodman, Rosing, Roucelle, Rumore, Russo, Saha,
  Sassolas, Schalk, Schellart, Schindler, Schmidt, Schneider, Schneider,
  Schoening, Schumacher, Schwamb, Sebag, Selvy, Sembroski, Seppala, Serio,
  Serrano, Shaw, Shipsey, Sick, Silvestri, Slater, Smith, Smith, Sobhani,
  Soldahl, Storrie-Lombardi, Stover, Strauss, Street, Stubbs, Sullivan,
  Sweeney, Swinbank, Szalay, Takacs, Tether, Thaler, Thayer, Thomas, Thornton,
  Thukral, Tice, Trilling, Turri, Berg, Berk, Vetter, Virieux, Vucina, Wahl,
  Walkowicz, Walsh, Walter, Wang, Wang, Warner, Wiecha, Willman, Winters,
  Wittman, Wolff, Wood-Vasey, Wu, Xin, Yoachim, \& Zhan}]{Ivezic_2019}
Željko Ivezić, Kahn, S.~M., Tyson, J.~A., {et~al.} 2019, The Astrophysical
  Journal, 873, 111, \dodoi{10.3847/1538-4357/ab042c}

\end{thebibliography}

\appendix

\section{Estimation of the redshift distribution}
\label{sec:photozdetail}

From UNIONS $r$-band observations, we follow three steps to estimate the redshift distribution of our weak-lensing source sample.
The first step is assigning multi-band photometry to UNIONS galaxies. Using the overlap of UNIONS $r$-band observations with the CFHTLenS \cite[Canada-France-Hawaii Telescope Lensing Survey; ][]{, 2012MNRAS.427..146H,2013MNRAS.433.2545E} W3 field ($\sim 44.2$ sq. deg), we assign $ugriz$ magnitudes by cross-matching. This can be done since CFHTLenS has deeper photometry \citep{2012MNRAS.421.2355H} than UNIONS, basically all CFIS (UNIONS $r$-band) objects are also visible in CFHTLenS, and the underlying redshift distribution is assumed to be the same after matching.

\begin{figure*}[htb]
     \centering
    \gridline{\fig{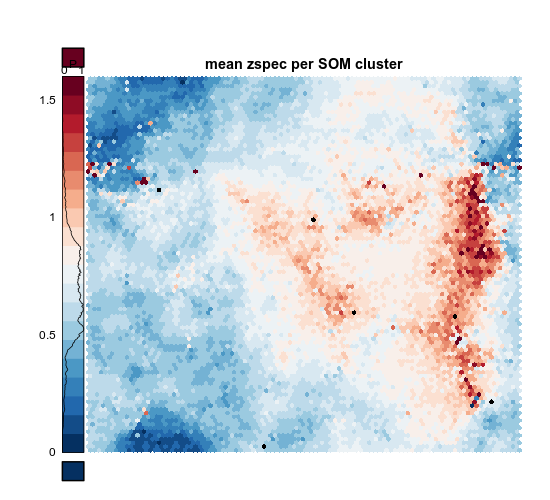}{0.5\textwidth}{(a) The SOM coloured in the spectroscopic sample}
          \fig{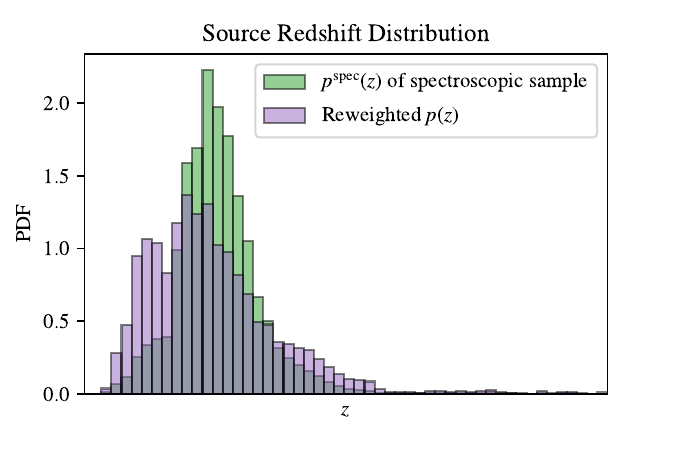}{0.5\textwidth}{(b) The blinded source redshift distributions}}
    \caption{In panel (a), the trained SOM coloured by
the counts of the spectroscopic sample is shown. In panel (b), the blinded redshift distributions $p^{\rm spec}(z)$ and $p(z)$ are plotted with green and purple bars, respectively.}
    \label{fig:photoz}
\end{figure*}

We calibrate the redshifts distribution with spectroscopic calibration samples which are constructed from DEEP2 \cite[DEEP2 Galaxy Redshift Survey; ][]{2013ApJS..208....5N}, VVDS \cite[VIMOS VLT Deep Survey; ][]{2005A&A...439..845L}, and VIPERS \cite[VIMOS Public Extragalactic Redshift Survey; ][]{2018A&A...609A..84S}. These surveys are also observed with CFHTLenS $ugriz$ photometry. With the multi-band photometry of the spectroscopic sample, we then train self-organising maps \cite[SOM; ][]{kohonen1982self,2015ApJ...813...53M} to organise the sample in high dimensional magnitude space. The SOM splits the matched sample into subsamples in its so-called SOM cells. The initial SOM cell grid has a resolution of 101 $\times$ 101 cells and is then hierarchically clustered into $5,000$ resolution elements for reliable statistics lateron, shown in Fig.~\ref{fig:photoz}. We then populate the SOM with the UNIONS weak lensing sources with $ugriz$ photometry. 

For every SOM cell $i$, a weight $w_{i}^{\mathrm{SOM}}$ is defined, which is the ratio of the number of UNIONS objects (weighted by their shape weights) over the number of spectroscopic objects
\citep{2020A&A...637A.100W}. Finally, we get the UNIONS $p(z)$ by re-weighting the spectroscopic redshift distribution $p^{\rm spec}(z)$ according to the weights $w_{i}^{\mathrm{SOM}}$ in the $i$'th SOM cells \citep{2020A&A...637A.100W},
\begin{equation}
    p(z)=\int w^{\mathrm{SOM}}(z)p^{\rm spec}(z) \mathrm{d} z=\sum w_{i}^{\mathrm{SOM}}p_{i}^{\mathrm{spec}}(z),
\end{equation}
where $p_{i}^{\mathrm{spec}}(z)$ is the histogram of spectroscopic objects per SOM cell $i$.
$p^{\rm Spec}(z)$ and $p(z)$ are shown in Fig.~\ref{fig:photoz}.

\section{Systematic Tests for Galaxy-galaxy Lensing Measurements}
\label{sec:systematic}

To validate our galaxy-galaxy lensing measurement, we conducted two null tests and measured the boost factor.

\subsection{Cross-shear ($\Delta \Sigma_\times$) test}

\begin{figure*}[bt]
    \centering
    \includegraphics[width=.85\columnwidth]{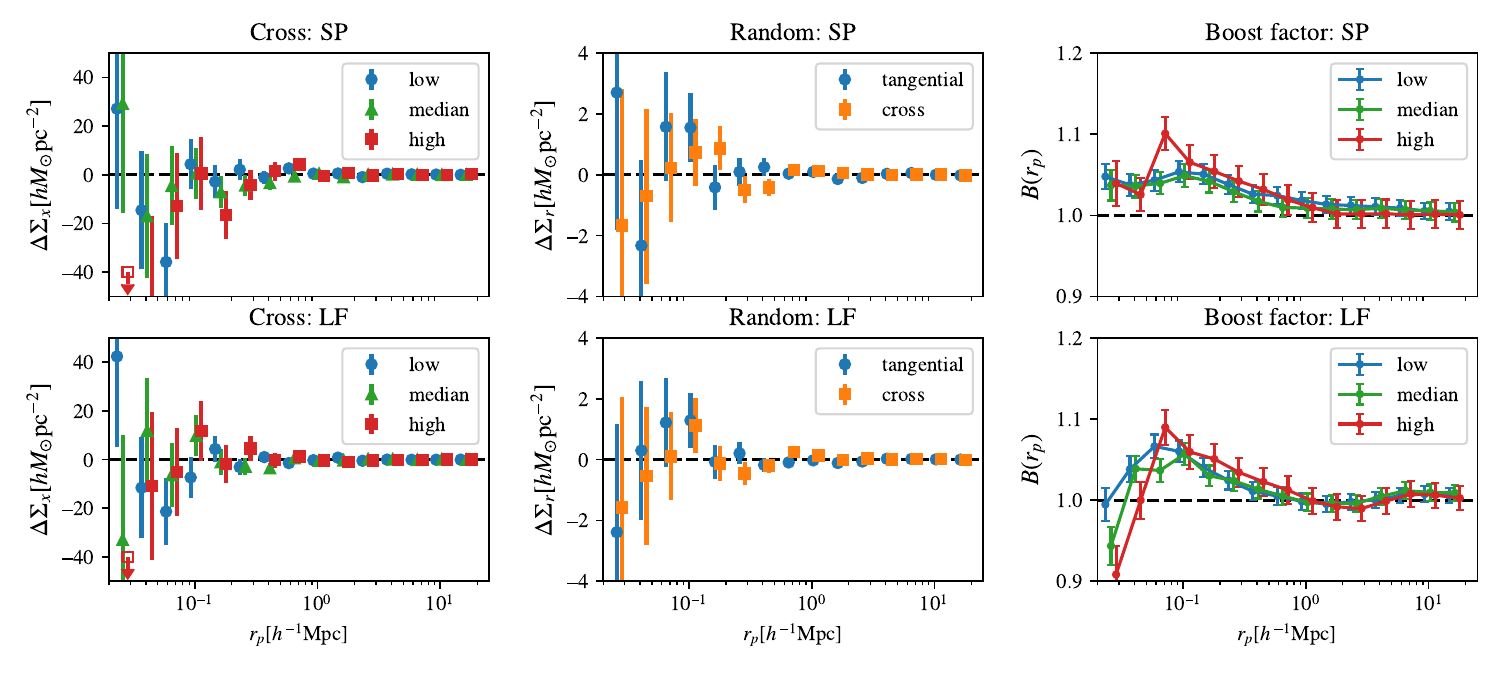}
    \caption{Systematic tests of lensing measurements. The top (bottom) column shows tests for ShapePipe (\textit{lens}fit). The left two panels correspond to $\Delta\Sigma_{\rm x}$. Low-, median-, and high-mass bins of type I sample are shown in blue circles, green triangles, and red rectangles in each panel. The middle two panels show the excess surface density and cross excess surface density around a random sample. The right two panels present the boost factor of the type I sample. Low-, median-, and high-mass bins are shown in blue, green, and red points. In all panels, some results are slightly displaced in the $x$-direction to make the figure clear.}
    \label{fig:ESDtest}
\end{figure*}

Weak gravitational lensing does not produce shape distortions in the cross direction, therefore the cross component of the shear $\gamma_\times$ or ``cross excess surface density'' $\Delta\Sigma_\times$ is expected to be zero in the absence of systematics. Thus, $\Delta\Sigma_\times$ can be interpreted as a null test of systematics in the lensing measurement process. We measure $\Delta\Sigma_\times$ with the same method and sample as for $\Delta\Sigma$,
\begin{align}
    \left\langle \Delta \Sigma(R)^w_\times \right\rangle
      & = \frac{\sum_{\rm l s} w_{\rm l} \,
        \overline{\Sigma_\textrm{cr}^{-1}}(z_{\rm l})
        \, w_{\rm s} \, \epsilon_{\times, {\rm s}}
                1_{b(R)}(| \vec r_{\rm l} - \vec r_{\rm s}|)
                }{ \sum_{\rm l s} w_{\rm l} w_{\rm s} \left( \overline{\Sigma_\textrm{cr}^{-1}}(z_{\rm l}) \right)^2 } .
\end{align} 
The results of the $\Delta\Sigma_\times$ test are shown in Fig.~\ref{fig:ESDtest}. All data points are consistent with zero at three sigma, and $\sim 70\%$ are zero within one sigma. No evidence is found of any significant systematic errors.

\subsection{Random lens test}

We also measure lensing signals around a random sample as a null test. This sample is constructed by randomly sampling the SDSS footprint and then selecting the sub-sample in the sky region overlapping with UNIONS. To match the redshift distribution, we randomly assign the redshifts of the high-mass bin lens sample (the other two bins have the same $p(z)$ after weighting) to the random sample. Our random sample contains $543,402$ ``galaxies''.

Following the same procedure as before, we measure both tangential and cross components with respect to the random sample with the ShapePipe and \textit{lens}fit shape catalogs. The results are presented in Fig.~\ref{fig:ESDtest}. The lensing signals are in good consistency with zero, indicating that systematic errors in the measurement are not significant.

\subsection{Boost factor}

Galaxy-galaxy lensing signals are diluted by galaxies physically associated with lens galaxies, whose shapes are not affected by lensing. Since we can not exclude these galaxies without photo-$z$s in this work, it is important to quantify this effect. With the same random sample as in the random lens test, we calculate the boost factor \citep{2004MNRAS.353..529H}, which is defined as $B(r)=(N_{\rm rs}\sum_{\rm ls}w_{\rm ls})/(N_{\rm ls}\sum_{\rm rs}w_{\rm rs})$, where $N_{\rm ls}$ and $N_{\rm rs}$ are the number of lens-source pairs and random-source pairs, respectively, and $w_{\rm ls}$ and $w_{\rm rs}$ are corresponding lensing weights. The results are shown in Fig.~\ref{fig:ESDtest}. We apply the boost factor correction to the lensing signals we use in this work.

\section{Details of the HOD model}
\label{sec:modeldetail}

\subsection{Baryonic contribution
}

For source-lens separations (at the lens redshifts) that are much larger than the size of a typical galaxy, that 
galaxy can be considered as a point mass. The baryonic contribution to the excess surface density, which contains stars, dust and gas, can then be written as
\begin{equation}
    \Delta\Sigma_\textrm{b}(R) = \frac{M_{*} }{\pi R^2} .
\end{equation}

\subsection{One-halo central galaxy contribution}

We adopt a Navarro-Frenk-White (NFW) model to describe the density profile of the host halo for central galaxies,
\begin{equation}
    \rho(r)=\frac{\rho_0}{(r/r_\textrm{s})(1+r/r_\textrm{s})^2},
\end{equation} 
with $\rho_0 = 200\,\rho_\textrm{m}/ (3 I_c)$ and $I_c = c^{-3} \int_0^c \textrm{d} x \, x (1+x)^{-2}$. Here, $\rho_\textrm{m}$ is the mean density of the Universe, and $c$ is the halo concentration parameter, defined as the ratio between the virial radius $r_{200}$ and scale radius $r_\textrm{s}$ of the halo, $c=r_{200}/r_\textrm{s}$.
Assuming that the halo center is located at the central galaxy, the excess surface density $\Delta\Sigma_{\rm NFW}$ of the halo within a disk of radius $R$ is
\begin{equation}
\label{eq:nfw}
    \Delta\Sigma_\textrm{h,cen}(R)   =\Delta\Sigma_{\rm NFW}(R)=\bar{\Sigma}_{\rm NFW}(<R)-\Sigma_{\rm NFW}(R)= \frac{M_{\rm h}}{2\pi r_\textrm{s}^2I}\left[ g(R/r_\textrm{s})-f(R/r_\textrm{s}) \right] ,
\end{equation}
where $M_{\rm h}$ is the halo mass. The functions $f(x)$ and $g(x)$ are defined as
\begin{equation}  
f(x)=
\left\{  
  \begin{array}{lr}  
  \frac{1}{x^2-2} \left[1-\frac{\ln[(1+\sqrt{1-x^2})/x]}{\sqrt{1-x^2}} \right], & x<1  \\  
  \frac{1}{3}, & x=1\\ 
  \frac{1}{x^2-1}\left[1-\frac{\arctan(\sqrt{x^2-1})}{\sqrt{x^2-1}}\right]; & x>1
  \end{array} 
\right. ,\qquad g(x)=  \left\{  
  \begin{array}{lr}  
  \frac{2}{x^2} \left[\ln{x/2}+\frac{\ln[(1+\sqrt{1-x^2})/x]}{\sqrt{1-x^2}} \right], & x<1  \\  
  2-2\ln 2, & x=1\\  
  \frac{2}{x^2} \left[\ln{x/2}+\frac{\arctan(\sqrt{x^2-1})}{\sqrt{x^2-1}} \right], & x>1 
  \end{array}  
\right. . 
\label{eq:fx}
\end{equation}

\subsection{One-halo satellite galaxy contribution}

We use the NFW model also for the host halo of satellites. Compared to the central-galaxy term, the satellite halo has a spatial offset. First, the excesses surface density given the projected distance between the satellite galaxy and the halo center, $R_\textrm{sat}$, is
\begin{equation}
\label{eq:off}
    \Delta\Sigma_{\rm off}(R|M_{\rm h, sat},R_{\rm sat}) = \frac{1}{2\pi}\int_0^{2\pi}\Delta\Sigma_{\rm NFW} \left( (R^2+R_{\rm sat}^2+2RR_{\rm sat}\cos\theta)^{\frac{1}{2}}|M_{\rm h, sat}\right) \textrm{d} \theta.
\end{equation}
We integrate this equation over the distribution functions of $M_{\rm h, sat}$ and $R_{\rm sat}$ to obtain the effective one-halo satellite term as
\begin{equation}
        \Delta\Sigma_{\rm h,sat}(R) = \iint \Delta\Sigma_{\rm off}(R|M_{\rm h, sat},R_{\rm sat}) P(R_{\rm sat}|M_{\rm h, sat}) P(M_{\rm h, sat}) \textrm{d} R_{\rm sat} \textrm{d} M_{\rm h, sat}
\end{equation}
We assume that satellite galaxies follow the spatial distribution of dark matter, which is the NFW density profile. We set
\begin{equation}
    P(R_{\rm sat}|M_{\rm h,sat})\propto \Sigma_{\rm NFW}(R_{\rm sat},M_{\rm h,sat})\cdot 2\pi R_{\rm sat} \textrm{d} R_{\rm sat}=\frac{M_{\rm h,sat}}{2\pi r_\textrm{s}^2(M_{\rm h,sat})I}f[R_{\rm sat}/r_\textrm{s}(M_{\rm h,sat}) ]\cdot 2\pi R_{\rm sat} \textrm{d} R_{\rm sat} .
\end{equation}
Following \cite{GS02}, we use a halo occupation distribution (HOD) model to infer
\begin{equation}
    P(M_{\rm h,sat}|M_{\rm sub})    \propto  P(M_{\rm sub}|M_{\rm h,sat})P(M_{\rm h,sat})    \propto  \langle N_{\rm sat}(M_{\rm h,sat})\rangle F_{\rm h}(M_{\rm h,sat}),
\end{equation}
where $F_{\rm h}(M_{\rm h})$ is halo mass function and $\langle N_{\rm sat}(M_{\rm h})\rangle$ is the halo occupation function of satellite galaxies. In this work, we use the halo mass function from \cite{TinkerHMF} and the HOD model from \cite{GS02}.

\subsubsection{Two-halo term}

For the two-halo term, we use the \cite{Tinker2010} halo bias model to infer the halo-matter correlation function $\xi_{\rm hm}=b_{\rm h}\xi_{\rm mm}$ based on the dark-matter correlation function $\xi_{\rm mm}$. From that, we can calculate the surface density as
\begin{equation}        
\Sigma_{\rm 2h}(R)=b_{\rm h}(M_{\rm h, cen}) \times 2\bar{\rho}\int_R^{\infty}\xi_{\rm mm}(r)\frac{r \textrm{d} r}{\sqrt{r^2-R^2}} .
\end{equation}

\subsection{Lens model validation}

To validate the model, we cross-match the type II sample with the \cite{2007ApJ...671..153Y} SDSS group catalog to select a purely central-galaxy subsample. Subsequently, we measure the ESD for both the type II sample and the central-galaxy subsample using the ShapePipe catalog. Next, we fit the central-galaxy subsample lensing signals with the lens model (Sect.~\ref{sec:model}), but set the satellite fraction to zero. This allows us to make two consistency tests. First, we compare the contributions of central galaxies from the entire type II sample by using the best-fit central ESD term to the ESD measured from the central-galaxy subsample. We find that they are broadly consistent. Second, we compare the inferred halo masses. They are consistent within one sigma across all mass ranges considered in this work, indicating that our lens model is reliable. 

\end{document}